\documentclass{aa}

\usepackage{graphicx}
\usepackage{txfonts}
\usepackage{natbib}
\usepackage[breaklinks=true, colorlinks=true, allcolors=blue]{hyperref}
\usepackage{amsmath}	
\usepackage{amssymb}	
\usepackage{multirow}
\usepackage{longtable}
\usepackage{threeparttable}
\usepackage{color}
\usepackage{rotating}
\usepackage{cancel,soul,ulem,amsmath} 
\usepackage{amsfonts}
\usepackage{amsbsy}
\usepackage{makecell}
\usepackage{gensymb}
\usepackage{float}
\usepackage{enumerate}
\usepackage{lastpage}
\usepackage{natbib}
\usepackage{array}
\usepackage{lscape}
\usepackage{tablefootnote}
\usepackage[justification=centering]{caption}
\usepackage{multicol}
\usepackage{graphicx}
\usepackage{epstopdf}
\usepackage{subfigure}
\usepackage{appendix}
\usepackage{lineno}
\usepackage{booktabs}

\begin{document}

   \title{Observational studies on  S-bearing molecules in massive star forming regions}


   \author{R. Luo\inst{1},
          J. Z. Wang\inst{1},
          X. Zhang\inst{2},
          D. H. Quan\inst{3},
          X. J. Jiang\inst{3},
          J. Li\inst{4},
          Q. Gou\inst{5}, 
          Y.Q. Li\inst{4},
          Y.N. Xu\inst{1},
          S.Q. Zheng\inst{4},
          C. Ou\inst{1},
          \and
          Y.J. Liu\inst{6}
          }

   \institute{Guangxi Key Laboratory for Relativistic Astrophysics, School of Physical Science and Technology, Guangxi University, Nanning 530004,  PR China\\
              \email{junzhiwang@gxu.edu.cn}
         \and 
             Xinjiang Astronomical Observatory, Chinese Academy of Sciences, 150 Science 1-Street, Urumqi, Xinjiang 830011, PR China
         \and
             Research Center for Intelligent Computing Platforms, Zhejiang Laboratory, Hangzhou, 311100, PR China
             \and
             Shanghai Astronomical Observatory, Chinese Academy of Sciences No. 80 Nandan Road
Shanghai, 200030, PR China
                \and
                School of Chemistry and Chemical Engineering, Chongqing University, Daxuecheng South Road. 55, Chongqing 401331, PR China
                \and
             Department of Physics, Anhui Normal University, Wuhu, Anhui 241002, PR China
             }

 \date{Received xx; accepted xxx}
 \authorrunning{Luo et al.}


\abstract
   {S-bearing molecules are powerful tools for  physical conditions inside massive star forming region. The abundances of S-bearing molecules, including H$_{2}$S, H$_{2}$CS, and HCS$^{+}$, are highly dependent on physical and chemical changes, which means that they are good tracers of the evolutionary stage of massive star formation. }
   {We present  observational results of  H$_{2}$S 1$_{10}$-1$_{01}$, H$_{2}$$^{34}$S 1$_{10}$-1$_{01}$, H$_{2}$CS 5$_{14}$-4$_{14}$, HCS$^{+}$ 4-3, SiO 4-3, HC$_{3}$N 19-18 and C$^{18}$O 1-0
toward  a sample of  51 late-stage massive star-forming regions, to study  relationships among H$_{2}$S, H$_{2}$CS, HCS$^{+}$ and SiO in hot cores. Chemical connections of these S-bearing molecules are discussed based on the relations between relative abundances in sources.}
   {H$_{2}$$^{34}$S 1$_{10}$-1$_{01}$, as the isotopic line of   H$_{2}$S 1$_{10}$-1$_{01}$, was used to correct  optical depths of H$_{2}$S 1$_{10}$-1$_{01}$. 
   Beam averaged  column densities of all molecules were calculated, as well as relative  abundances of  H$_{2}$S, H$_{2}$CS, and HCS$^{+}$  to that of H$_{2}$, which were derived from  C$^{18}$O. One chemical model  including  gas,  dust grain surface and  icy mantle phases, for H$_2$S, H$_2$CS, and HCS$^+$ molecules, was used to compare with the observed abundances.}
   {H$_{2}$S 1$_{10}$-1$_{01}$,  H$_{2}$$^{34}$S 1$_{10}$-1$_{01}$, H$_{2}$CS 5$_{14}$-4$_{14}$, HCS$^{+}$ 4-3 and HC$_{3}$N 19-18 were detected in 50 of the 51 sources, while SiO 4-3 was detected in 46 sources.  C$^{18}$O 1-0  was detected in all sources.  
   The Pearson correlation coefficients between H$_{2}$CS and HCS$^+$ normalized by H$_{2}$ and H$_{2}$S are 0.94 and 0.87, respectively,  and  a tight linear relationship is found between them with slope of 1.00 and 1.09,   while  they are  0.77 and 0.98  between H$_{2}$S and  H$_{2}$CS,  respectively,  and   0.76 and 0.97 between H$_{2}$S and HCS$^{+}$.  The values of  full width at half maxima (FWHM)   of H$_{2}$$^{34}$S 1$_{10}$-1$_{01}$, H$_{2}$CS 5$_{14}$-4$_{14}$, HCS$^{+}$ 4-3, and HC$_{3}$N 19-18 in each source are similar to each other, which indicate that they can trace  similar regions. Comparing the observed abundance with model results, there is one possible time (2-3$\times$10$^{5}$ yr) for each source in the model that match the measured abundances of H$_{2}$S, H$_{2}$CS and HCS$^{+}$. The abundances of HCS$^{+}$, H$_{2}$CS, and H$_{2}$S increase with the increment of SiO abundance in these sources, which implies that shock chemistry may be important for them. }   
  {Close abundance relation of H$_2$S,  H$_2$CS and HCS$^+$ molecules and   similar line widths in observational results indicate that these three molecules could be chemically linked, with HCS$^+$ and H$_2$CS the most correlated. The comparison of the observational results with chemical models shows that the abundances can be reproduced for almost all the sources at a specific time. The observational results,  including abundances in these sources need to be considered in further modeling H$_{2}$S, H$_{2}$CS and HCS$^{+}$ in hot cores with shock chemistry.}
   {}


\maketitle
\nolinenumbers

%

\section{Introduction}

As the 10th most abundant element in the universe with relative abundance to hydrogen of  $\sim$ 1.3×10$^{-5}$ \citep{asp05},  sulfur (S) was found in many molecules, which is important for understanding gas chemistry in molecular clouds with different phases.  
There are  more than a dozen S-bearing molecules that were  detected in interstellar molecular clouds and/or outflows, including  SH, SH$^{+}$, SO, SO$_{2}$, CS, C$_{2}$S, C$_{3}$S, CH$_{3}$SH, NS, SiS, H$_{2}$S, H$_{2}$CS, HCS$^{+}$, OCS,  HNCS, and HSCN   \citep{go78, fre79, br80, bla87, drd89, tu89, ad10, esp13, neu15, li15, luo19, ce20}.  Even though  S abundance  in diffuse HI medium  is  still consistent with cosmic value \citep {jen09}, the main reservoir of S in dense clouds and hot core is still unclear \citep{vidal18}, since the derived abundance of dense clouds only account for $\sim$ 0.01 of the cosmic abundance of atomic S \citep{cha97}.

S-bearing molecules are thought to be  useful tracers of  chemical and physical properties of complex star forming regions (SFRs) located in dense molecular clouds and can be used to study the evolutionary  stages of massive star formation \citep{esp14}.  \citet{esp13} showed that S-bearing molecules exhibit higher column densities, up to three orders of magnitude,  in the high-velocity plateau of Orion KL affected by shocks, with respect to quiet regions of the clouds. Therefore, they are also considered to be good  tracers for shocks. S-bearing molecules, released  from  ice mantles  at T $\sim$ 100-300 K, such as SO, H$_{2}$S, H$_{2}$CS, OCS, can probe hot cores in low-mass protostellar Class 0 system \citep {ty21}. \citet{fo23} find for a sample of massive star-forming regions that NS, CCS, and HCS$^{+}$ trace cold, quiescent, and likely extended material, while OCS and SO$_{2}$ trace warmer, more turbulent, and likely denser and more compact material. On the other hand,  SO traces both quiescent and turbulent material. But the chemical origin  of  H$_2$S and  H$_2$CS is less clear.



Sulfur and oxygen are part of the chalcogens group in the periodic table, with  similar  electronic structure and chemical bond,  meaning that they usually have a similar reactivity in chemical reactions \citep{mi13}.  As the third most abundant element  after hydrogen and helium in the interstellar medium (ISM), oxygen cosmic abundance is of O/H $\sim$ 2.5×10$^{-4}$ \citep{lis23}. H$_2$O, H$_2$CO and HCO$^+$, listed in order of decreasing abundances in the ISM,  are important oxygen-containing molecules with close astrochemical relationships. Replacing O by S,  H$_2$S, H$_2$CS and HCS$^+$ molecules  should play important roles for understanding chemical network of S-bearing molecules in ISM, with relative high abundances among  S-bearing molecules. For example, the abundances of  H$_2$S, H$_2$CS and HCS$^+$ molecules  normalized by H$_{2}$ are 4.6$\times$10$^{-8}$, 1.3$\times$10$^{-9}$ and 1.5$\times$10$^{-10}$, respectively, in G328.2551-0.5321 \citep {bou22}.

There are few observational studies for H$_2$S, H$_2$CS, and HCS$^+$ molecules toward  hot cores  \citep {esp13, li15, ka22, el22, esp23}.  {The dominant S-bearing species in ices are still in debate. H$_{2}$S was suggested as one, since predicted to be to abundant on the grain surface in low gas and dust temperature environment \citep [$<$20 K] {vidal17, nav20}, but it has not still been detected on ices \citep{mcc23}. Once H$_2$S molecules formed, they rapidly initiate reactions that drive the production of other S-bearing molecules \citep{wak12}. 
The formation of S-bearing molecules is closely related to the environment, i.e. temperature \citep{wak12}, UV photons and high energy protons  \citep{jim11},  ion irradiation \citep {gar10}. In molecular clouds, ion-molecular reactions are most important  \citep{gol80}, while in warm gases of hot cores and shock waves, neutral-neutral reactions mainly occur  \citep{cha97}.  H$_{2}$CS may be formed by CH + H$_{2}$S in the gas phase of massive star forming regions such as Orion KL and Sgr B2 as predicted by  a astrochemical model \citep{in20}. What's more, H$_{2}$CS can be formed via S + CH$_{3}$ and destroyed with C$^{+}$ in gas-phase \citep{esp22}. Finally, HCS$^{+}$  may be formed by CS + H$_{3}$$^{+}$ and CS$^{+}$ + H$_{2}$ in high-mass star forming cores. \citep{shi20, fo23} The dissociative recombination  rate of HCS$^+$ is substantially larger than those of other isoelectronic ions, which had substantial effect on the abundances of molecules and ions in the low-temperature environment of interstellar clouds  \citep {mon05}. However, there are still lack of sufficient studies, including H$_{2}$S, H$_{2}$CS, HCS$^{+}$ detections and the relationship among the three molecules in massive star forming regions.

In this paper, we present H$_{2}$S 1$_{10}$-1$_{01}$, H$_{2}$$^{34}$S 1$_{10}$-1$_{01}$, H$_{2}$CS 5$_{14}$-4$_{14}$, HCS$^{+}$ 4-3, SiO 4-3, HC$_{3}$N 19-18 and C$^{18}$O 1-0 single pointing observations  toward a sample of 51 massive star forming regions  using the Institut de Radioastronomie Millim$\acute{\rm{e}}$trique (IRAM) 30-m telescope. The observations are described in Section \ref {sec:observation}, while the results are reported in Section \ref {sec:result}.  Discussions are presented in Section \ref {sec:dis}, and a brief summary is  in Section \ref {sec:conclusion}.

\begin{figure*}
 \centering 
\subfigure[]{ \label{fig1:a} 
 \includegraphics[width=0.65\columnwidth]{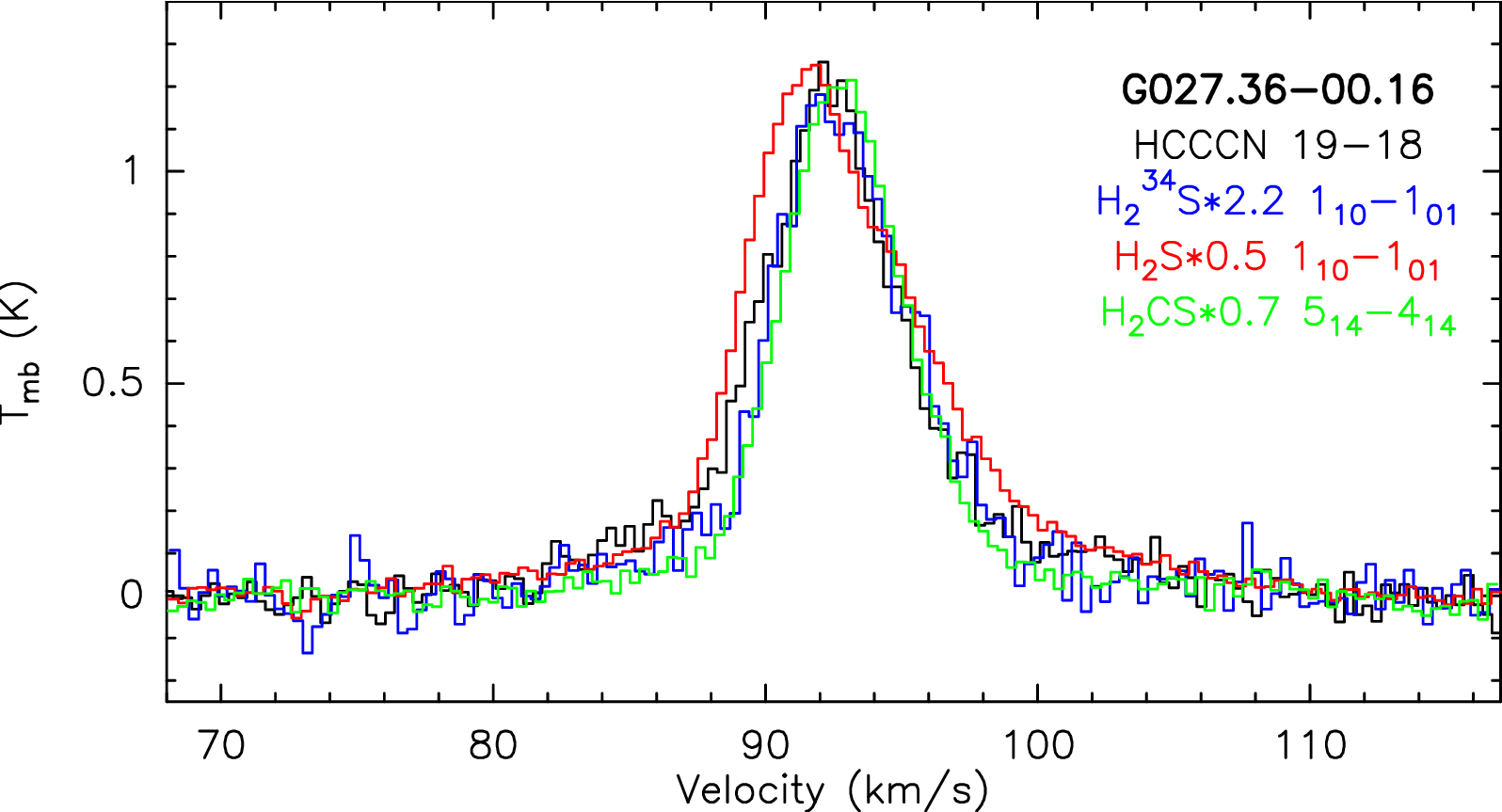} 
} 
\subfigure[]{ \label{fig1:b} 
\includegraphics[width=0.65\columnwidth]{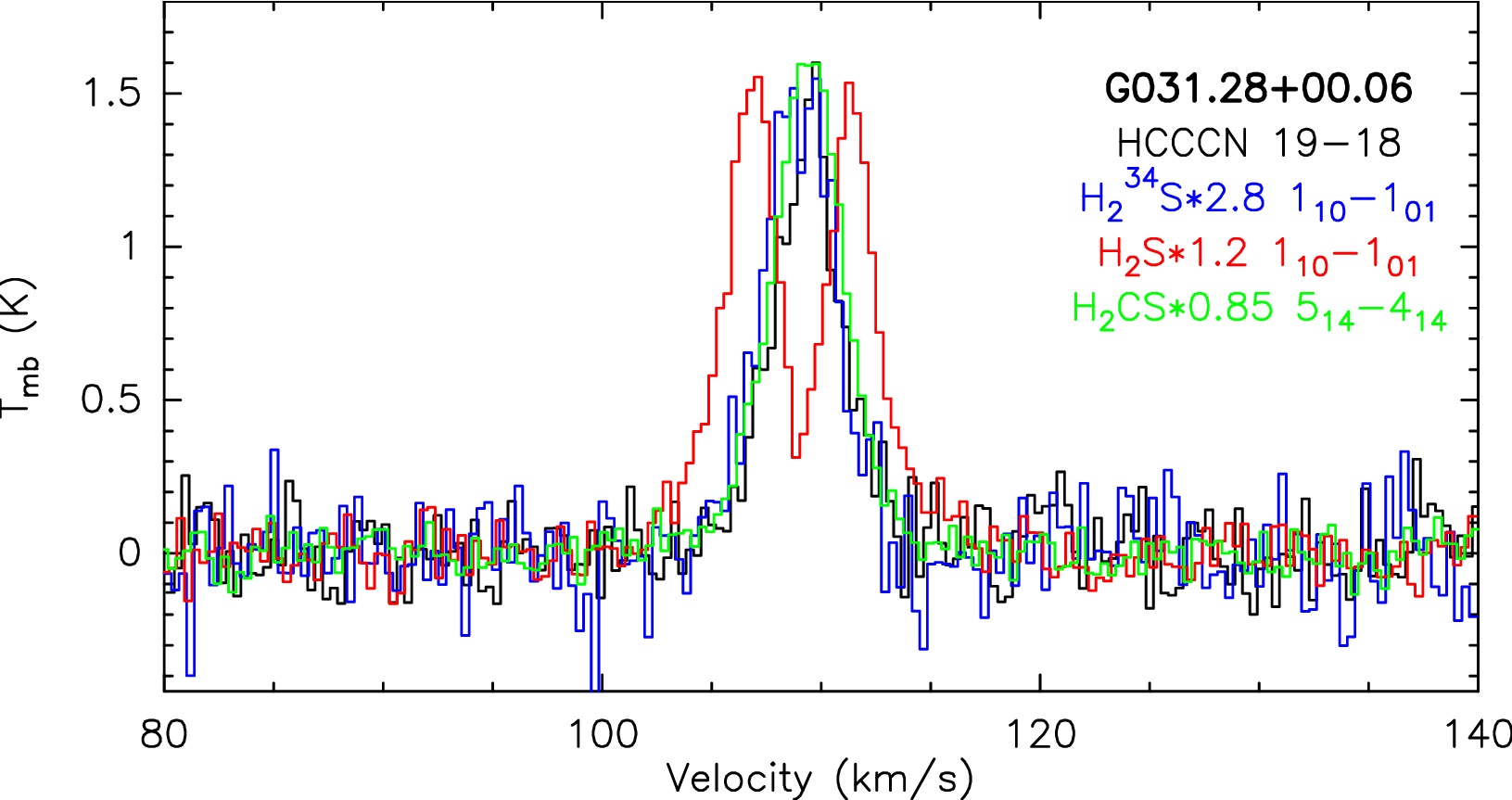}

} 
\subfigure[]{ \label{fig1:c} 
\includegraphics[width=0.65\columnwidth]{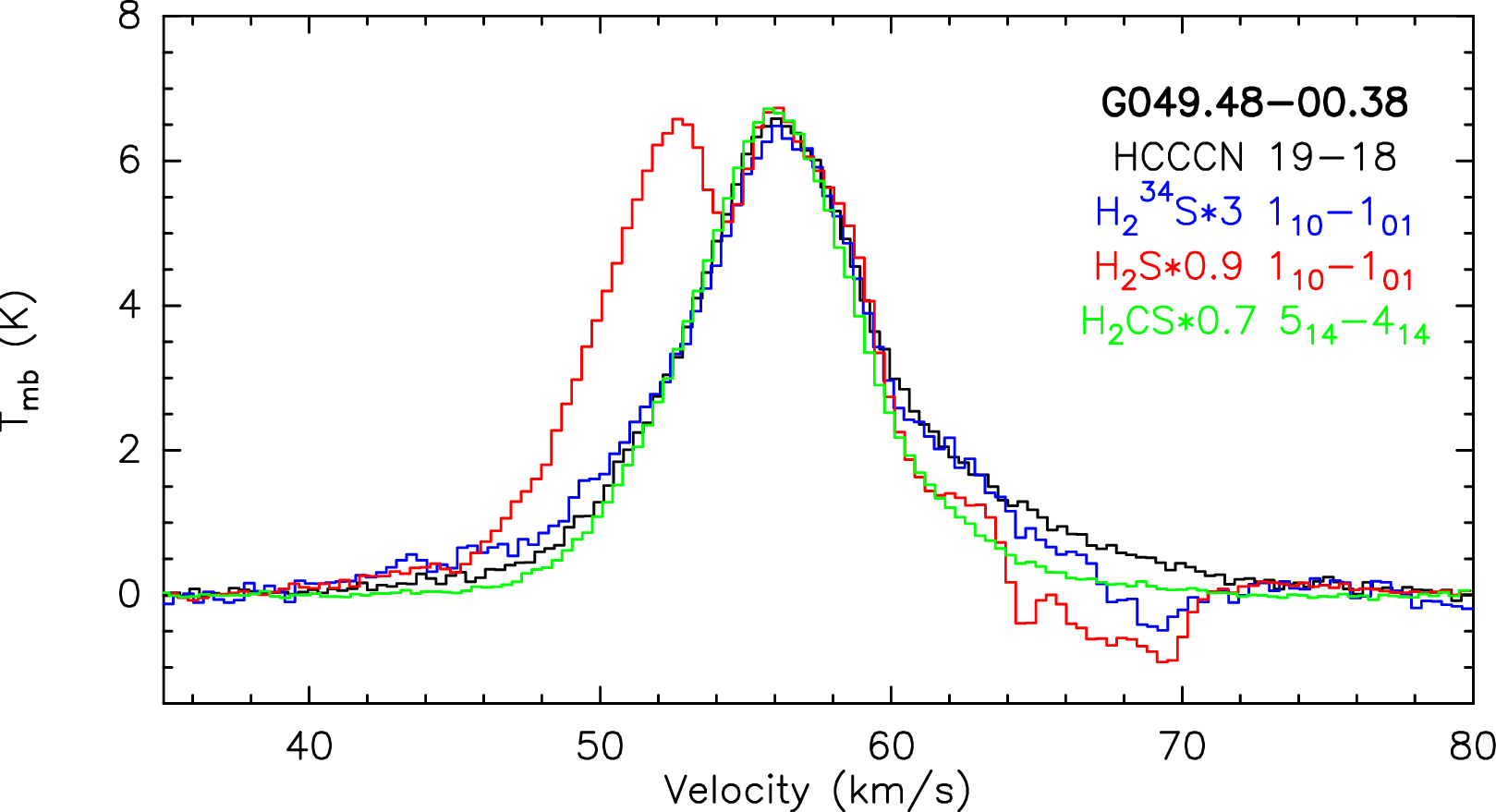}

} 
\subfigure[]{ \label{fig1:d} 
\includegraphics[width=0.65\columnwidth]{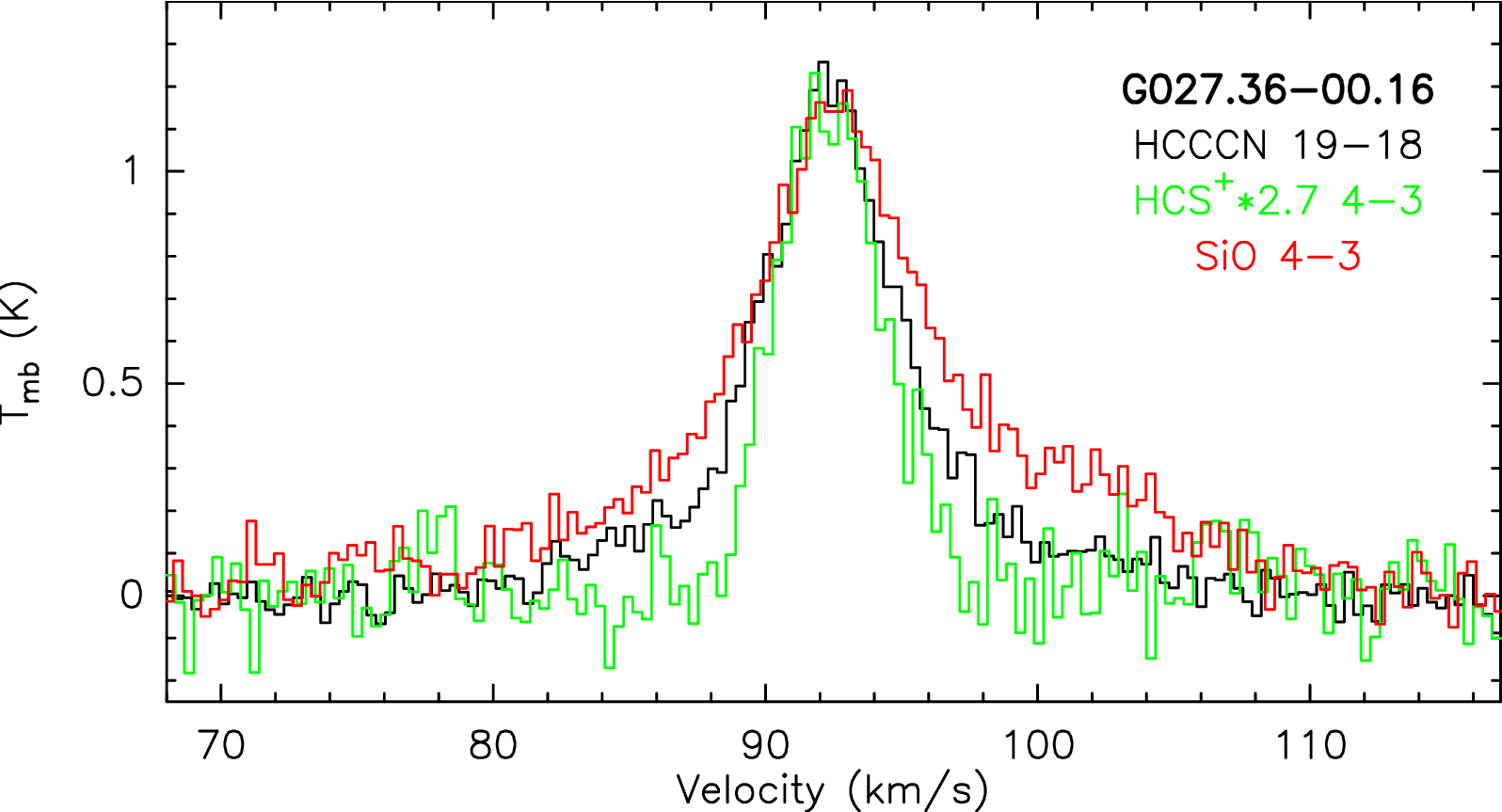}

}
\subfigure[]{ \label{fig1:e} 
\includegraphics[width=0.65\columnwidth]{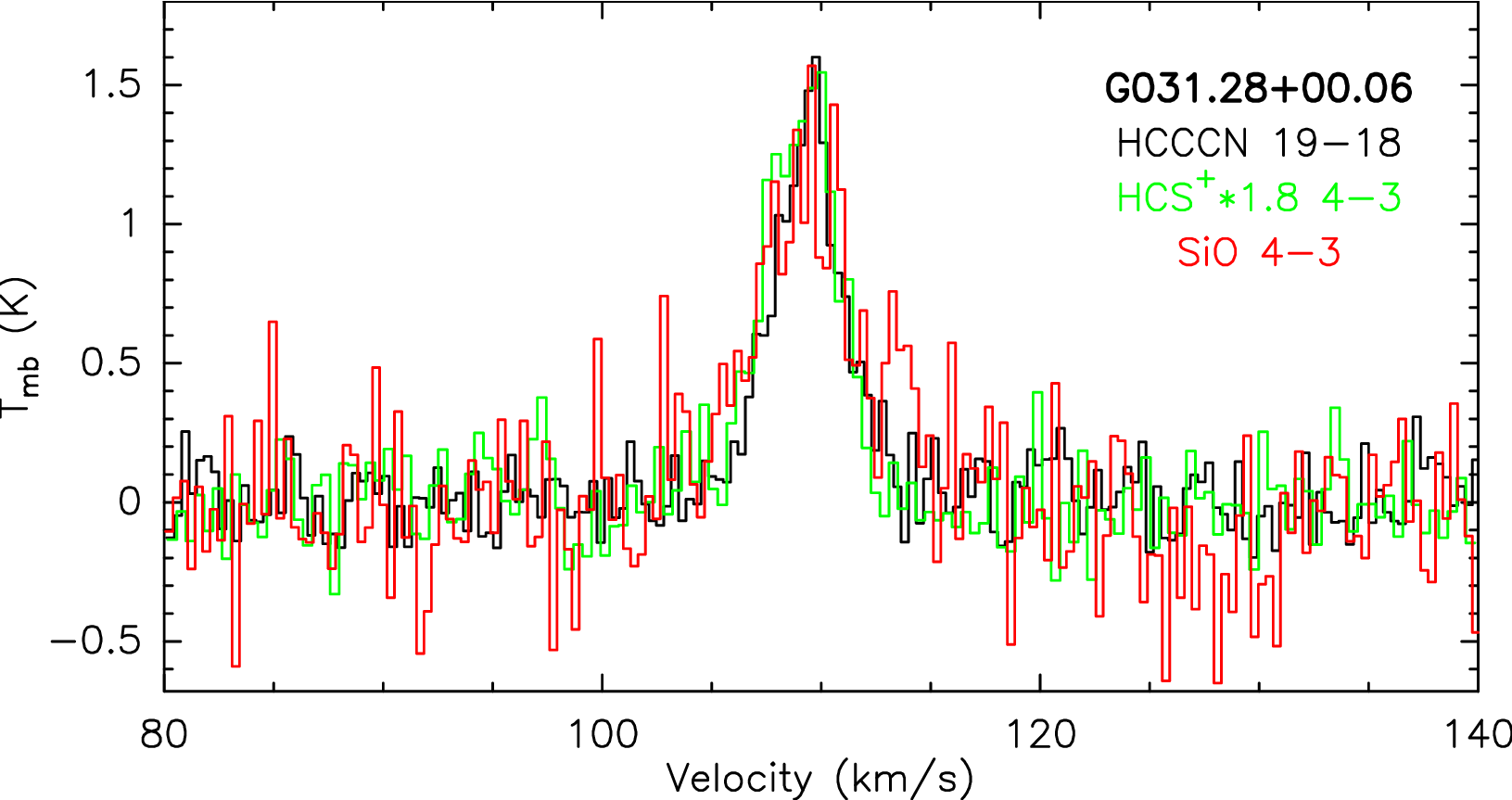}

} 
\subfigure[]{ \label{fig1:f} 
\includegraphics[width=0.65\columnwidth]{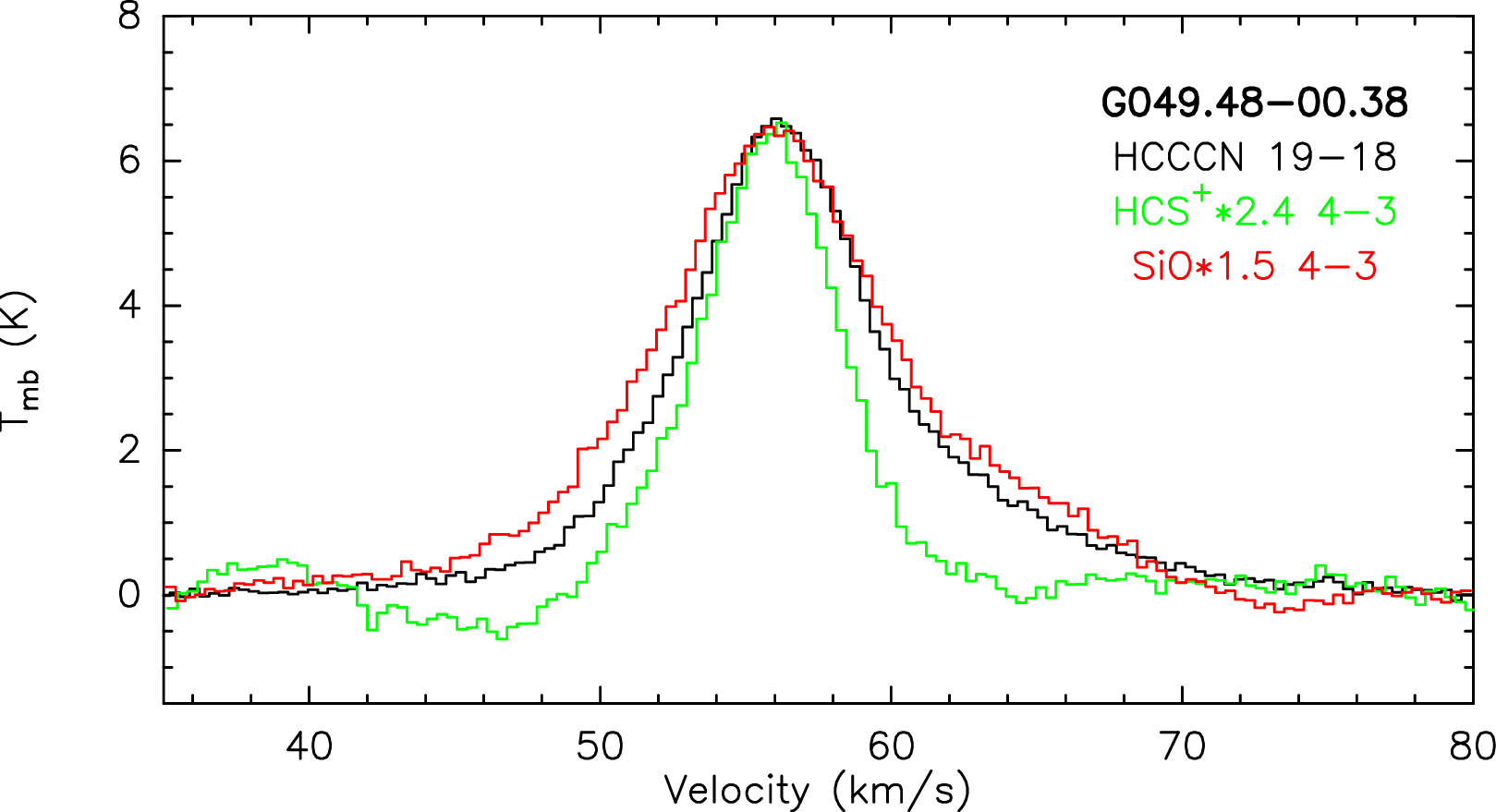}

}  
\caption{The spectra of  H$_{2}$S 1$_{10}$-1$_{01}$, H$_{2}$$^{34}$S 1$_{10}$-1$_{01}$, H$_{2}$CS 5$_{14}$-4$_{14}$ detected with IRAM 30-m for three sources are shown in (a), (b) and (c), while HCS$^{+}$ 4-3, SiO 4-3 lines are shown in (d), (e) and (f). All of the observed spectra are aligned to the peak with  HC$_{3}$N  19-18. In (a), (b) and (c), red lines are H$_{2}$S 1$_{10}$-1$_{01}$, blue lines are H$_{2}$$^{34}$S 1$_{10}$-1$_{01}$, green lines are H$_{2}$CS 5$_{14}$-4$_{14}$. In (d), (e) and (f), HCS$^{+}$ 4-3 lines are marked in green, while SiO 4-3 lines are red. The black lines in all figure are HC$_{3}$N}.
\label{fig:fig1}
\end{figure*}

\section{Observations and analysis}\label{sec:observation}
\subsection{Observations and data reduction} \label{obss}

The sample, which is selected from  \citet{reid14,reid19} with  measurements of trigonometric parallaxes, includes 51 late-stage massive star-forming regions with 6.7 GHz methanol masers.    The observations were taken  with Institut de Radioastronomie Millim$\acute{\rm{e}}$trique (IRAM)  30-m millimeter telescope at Pico Veleta Spain during 2016 June  and October, 2017 August, and 2020 August. 
The 3-mm (E0) and 2-mm (E1) bands of the Eight Mixer Receiver (EMIR) were used simultaneously with the Fourier Transform Spectrometers (FTS) backend to cover 8 $\times$ 4 GHz bandwidth and 195 kHz spectral resolution for each band with dual polarization. More information can be found in  \citet{li22}. The spectral lines, H$_{2}$S 1$_{10}$-1$_{01}$ at 168762.7623MHz, H$_{2}$$^{34}$S 1$_{10}$-1$_{01}$ at 167910.516MHz, H$_{2}$CS 5$_{14}$-4$_{14}$ at 169114.160MHz, HCS$^{+}$ 4-3 at 170691.603MHz, HC$_{3}$N 19-18 at 172849.287MHz, SiO 4-3 at 173688.274MHz, and C$^{18}$O 1-0 at 109782.1734MHz, were included in the observations and used for scientific analyses in this paper.  Table \ref {table:source} shows the source names, aliases, equatorial coordinates and galactocentric distance in columns 1-4.  The spectroscopic information parameters of H$_{2}$S,  H$_{2}$$^{34}$S,  H$_{2}$CS,  HCS$^{+}$, SiO, HC$_{3}$N and C$^{18}$O lines are presented in Table \ref{table:spectro}.

The beam size of IRAM 30-m telescope is 22.4$^{\prime\prime}$ at 110 GHz and 14.3$^{\prime\prime}$ at 172 GHz. The main beam brightness temperature ($T_{mb}$) is obtained by $T_{mb} = T_{A}^{\ast} \cdot F_{eff} / B_{eff}$,  where $T_{A}^{\ast}$ is antenna temperature, while $F_{eff}$ is forward efficiency and $B_{eff}$  is main beam efficiency. $F_{eff}$  and $B_{eff}$ are  94$\%$ and 78$\%$ in 3-mm band at 110 GHz, respectively, while they are  93$\%$ and  73$\%$, respectively, in 2-mm band at 173 GHz.

The data was proceeded with CLASS package in GILDAS software\footnote{\url{http://www.iram.fr/IRAMFR/GILDAS}}. After averaging  the spectra for each sources to one single spectrum at 2mm and 3mm band respectively,  the lines were identified and   first-order baseline was removed. The velocity integrated  flux, peak intensity, central velocity of each line were derived for most of the lines with single component Gaussian fitting.  Otherwise, "print area'' in CLASS was used to derive velocity integrated  flux if the line profile can not be well fitted with single component Gaussian fitting (see Figure \ref {fig:fig1}),  which are caused by  self-absorption and/or  multiple components (see Figure \ref {fig1:b}  and  \ref {fig1:c}), for lines above 3$\sigma$ level, while H$_{2}$$^{34}$S 1$_{10}$-1$_{01}$  was   used to determine the velocity range of each source. 

\subsection{Methods} \label{me}

An optically thin transition produces an antenna temperature that is proportional to the column density in the upper level of the transition being observed  \citep {gol99}. However, since sulfur is the 10th most abundant element in the universe and H$_{2}$S is likely the dominant molecular form  of the S element  in molecular clouds,  H$_{2}$S lines are usually optically thick. Some of the  H$_{2}$S 1$_{10}$-1$_{01}$ lines showed  a pit around line center on the spectrum level,  as the signature of self-absorption. Since H$_{2}$$^{34}$S 1$_{10}$-1$_{01}$ lines were also detected with  H$_{2}$S 1$_{10}$-1$_{01}$  detection, we use the beam averaged column densities of H$_{2}$$^{34}$S  to calculate those of H$_{2}$S, by $^{32}$S/$^{34}$S ratio, as function of galactocentric distance, which was from \citet {yan23} given by 

\begin{equation}
^{32}S/^{34}S = (0.75 \pm 0.13) R_{GC} + (15.52 \pm 0.78)
\end{equation}

\noindent The beam-averaged column density is given by:

\begin{equation}
N_{tot}=\frac{8\pi kv^{2}}{hc^{3}A_{ul}} \frac{Q(T_{ex})}{g_{u}} e^\frac{Eu}{kT_{ex}} \int T_{mb} \ {\rm d}v    \quad  \quad ({\rm cm}^{-2})
\end{equation}

\noindent  For above function, $k$ is the Boltzmann constant, $v$ is the transition frequency. $h$ is the Planck constant, $c$ is the speed of light, $A_{ul}$ is the Einstein emission coefficient, $g_{u}$ is upper state degeneracy  and $E_{u}$ is upper state energy.
For sources with non-detection, 3$\sigma$ upper limits for the integrated intensity, $\int T_{mb} \ {\rm d}v$, were calculated with $3\times rms \sqrt{\delta v \Delta v}  $, where $\delta$$v$ is the channel separation in velocity, 
$rms$ is the root mean square for line free channel from first-order linear fitting using the remaining channels, which are masked  the signal area, 
and $\Delta$$v$  is the line width.      

After assuming unity  beam filling factor, and  local thermodynamic equilibrium (LTE) approximation for all molecules with  the same excitation temperature ($T_{ex}$)  of  18.75 K for   H$_{2}$$^{34}$S,  H$_{2}$CS,  HCS$^{+}$, SiO, and C$^{18}$O, as well as optically thin for the lines of  all these molecules,  the beam averaged column densities were  calculated. Since the H$_{2}$$^{34}$S,  H$_{2}$CS,  HCS$^{+}$ molecules are normally sub-thermal in these sources, the used $T_{ex}$ of these spectral lines are lower than kinetic temperature ($T_{k}$). 
The partition function ($Q(T_{ex})$) for each molecule was obtained at $T_{ex}$=18.75 K,   checked   by Radex on line calculator \citep {van07}. 
For example, the upper level population of H$_{2}$$^{34}$S 1$_{10}$-1$_{01}$ over all energy levels (POP-UP) are obtained by RADEX. When T$_{k}$ ranges from 30 K to 150 K and density of H$_{2}$ is 1$\times$10$^{5}$ cm$^{-3}$, POP-UP increases from 0.163 to 0.267. $Q(T_{ex})$ were obtained from Cologne Database for Molecular Spectroscopy (CDMS) 
\footnote{\url{https://cdms.astro.uni-koeln.de/classic/predictions/catalog/partition_function.html}}   \citep{mu01, mu05}.  Under LTE approximation with $T_{ex}$ = 18.75 K, $g_{u}e^\frac{-Eu}{kT_{ex}}/Q(T_{ex})$, which is the same parameter as POP-UP in the results of RADEX for calculating total column density, is equal to 0.234.
We also calculated the systemic bias of estimating column densities with excitation temperature of 37.5 K (0.177), 25 K (0.222) and  10 K (0.150), which is about 1.32, 1.05 and 1.55 times, respectively, of that with $T_{ex}$ = 18.75 K for  H$_{2}$$^{34}$S  molecules.  The column densities of H$_{2}$ can be derived with C$^{18}$O  by $N_{\rm H_{2}} = 4.37 \times 10^{6} N_{\rm C^{18}O}$  \citep{fre82} .

In addition to using H$_{2}$$^{34}$S column densities with the ratio of $^{32}$S/$^{34}$S to directly calculate the column density of H$_{2}$S, there is another method that correcting the optical depths of H$_{2}$S to get the column density.
Assuming  H$_{2}$S and H$_{2}$$^{34}$S 1$_{10}$-1$_{01}$ lines in each source have  the same excitation temperature ($T_{ex}$) and molecular abundance ratio is approximately equal to isotopic  abundance ratio as $\frac{^{32}S}{^{34}S}=\frac{\rm{H}_{2}^{32}S}{\rm{H}_{2}^{34}S}$ (hear-after H$_2$S for H$_{2}$$^{32}$S). Then, $\tau_{H_{2}S}$ is calculated by 

\begin{equation}
\frac{1-e^{-\tau_{H_{2}S}}}{1-e^{-\tau_{H_{2}^{34}S}}}=\frac{\int T_{mb}(H_{2}S) \ {\rm d}v}{\int T_{mb}(H_{2}^{34}S) \ {\rm d}v}
\end{equation}

\noindent The range of calculated $\tau$$_{H_{2}S}$ is from 0.53 in G232.62+00.99 to 11.2 in G010.47+00.02 (see Table \ref {figA1}).
The calculation formula of column density is slightly different from that of Eq.(1), which given by 

\begin{equation}
N_{tot}=\frac{8\pi kv^{2}}{hc^{3}A_{ul}} \frac{Q(T_{ex})}{g_{u}} e^\frac{Eu}{kT_{ex}} \frac{\tau}{1-e^{-\tau}} \int T_{mb} \ {\rm d}v    \quad  \quad ({\rm cm}^{-2})
\end{equation}

Comparing the column densities obtained by the two methods, we found that the column densities directly from H$_{2}$$^{34}$S using the $^{32}$S/$^{34}$S ratio were lower than that from optical depth of the  H$_{2}$S transition, with 0.85 times in columns 1-2 of Table \ref {table:density}.

\begin{figure*}
 \centering 
\subfigure[]{ \label{fig2:a} 
 \includegraphics[width=0.65\columnwidth]{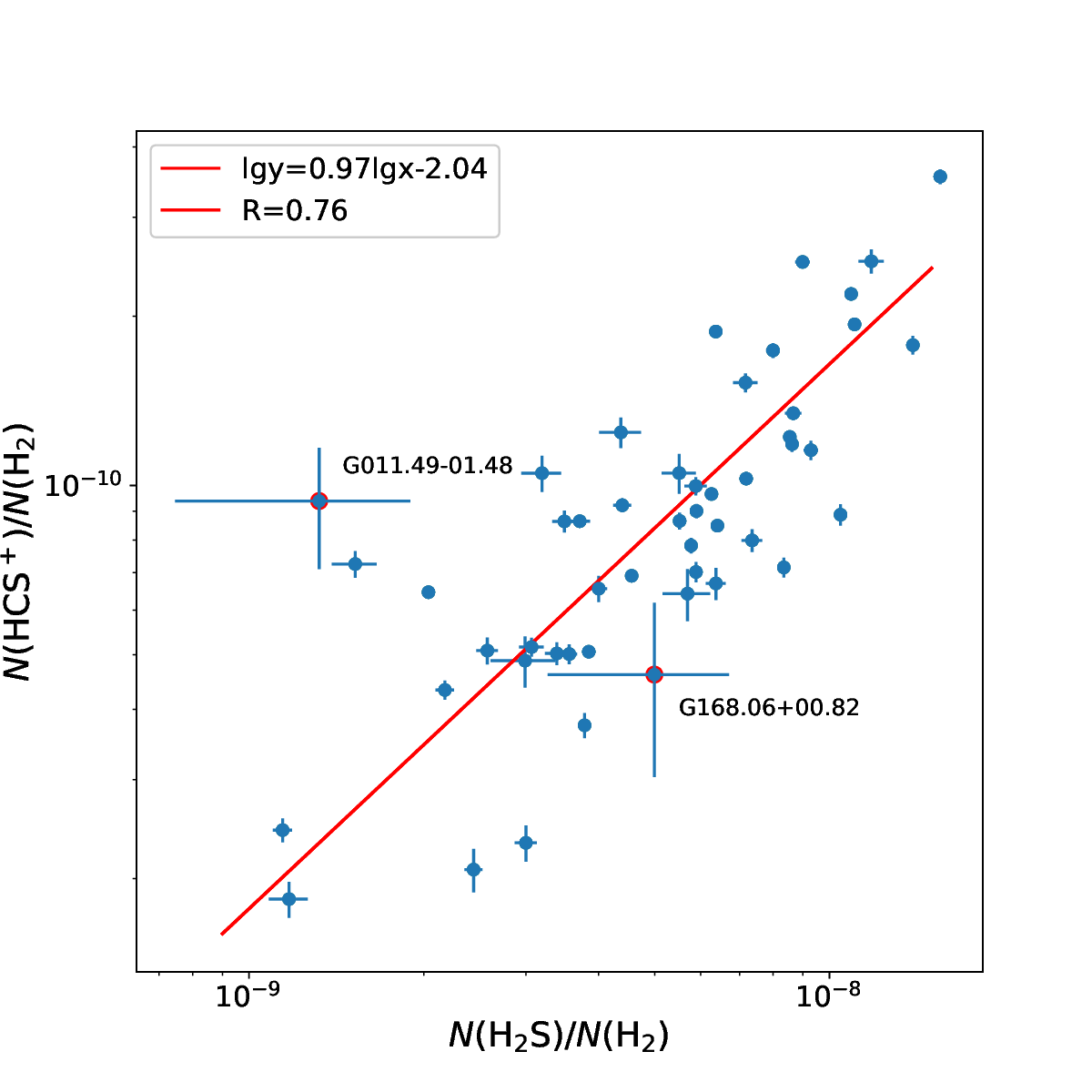} 
} 
\subfigure[]{ \label{fig2:b} 
\includegraphics[width=0.65\columnwidth]{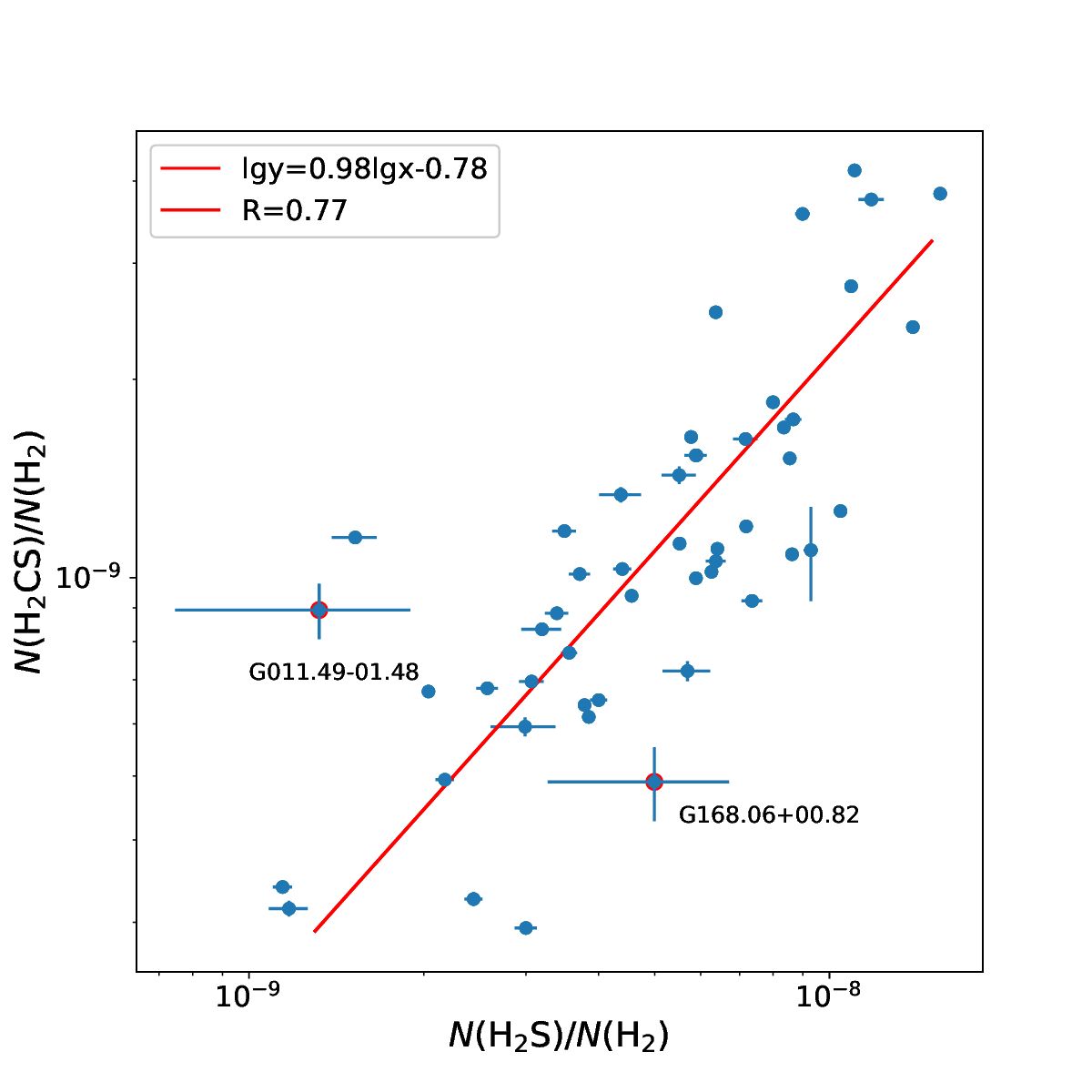}
} 
\subfigure[]{ \label{fig2:c} 
\includegraphics[width=0.65\columnwidth]{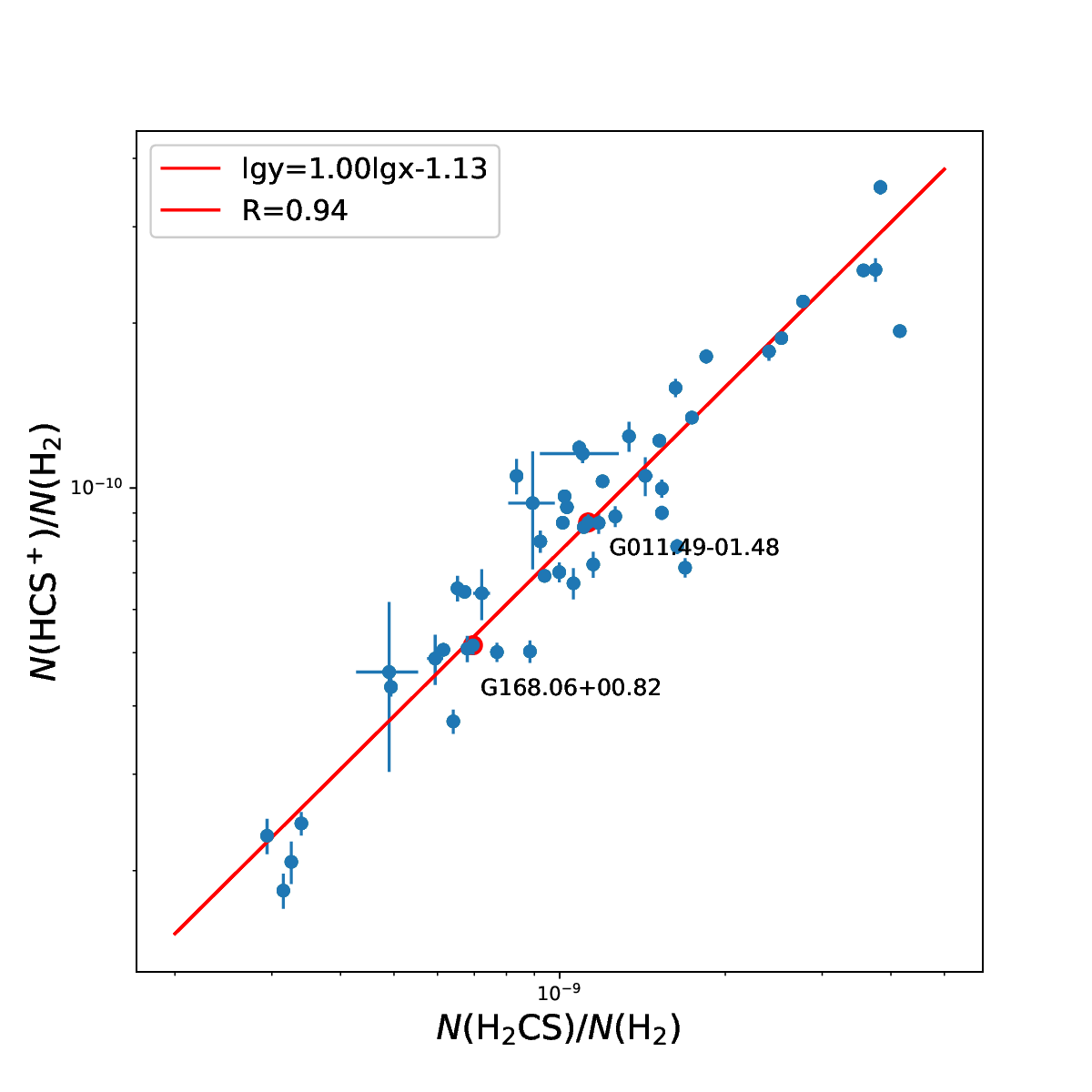}
} 
\caption{The relationship among H$_{2}$S, H$_{2}$CS and HCS$^{+}$ molecules normalized by H$_{2}$. (a) shows the relationship between HCS$^{+}$ and H$_{2}$S abundances. And (b) shows relationship between the HCS$^{+}$ and H$_{2}$S abundances, while (c) prints the relationship between HCS$^{+}$ and H$_{2}$CS abundances. 
The fitting lines are marked in red, using the least square method. G011.49-01.48 and G0168.06+00.82, which have large errors,   are also marked, where H$_{2}$$^{34}$S lines showed weak emission.}
\label{fig:fig2}
\end{figure*}

\begin{figure*}
 \centering 

\includegraphics[width=1\columnwidth]{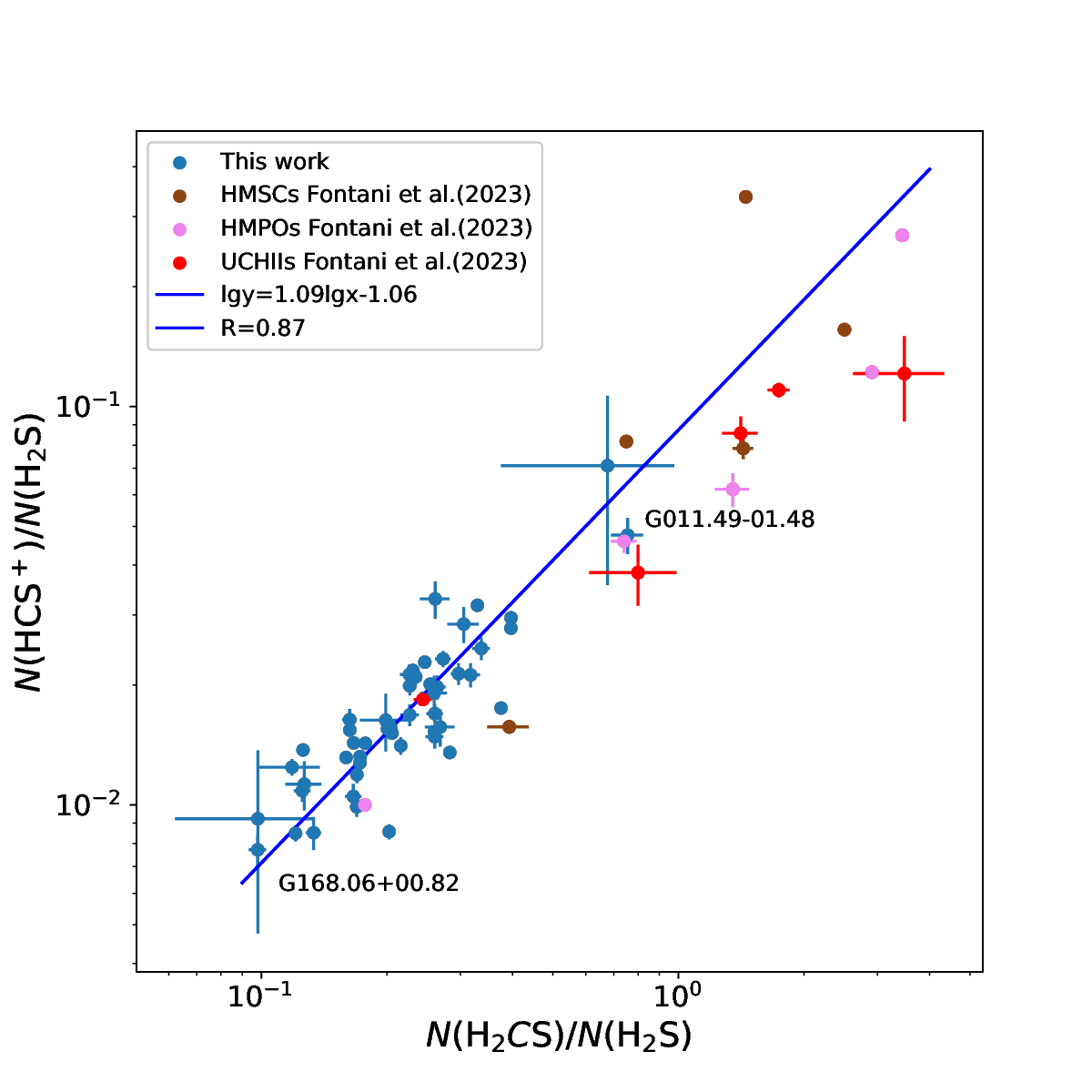}

\caption{The distribution of  H$_2$CS and HCS$^+$ molecular abundances in each source. Two sources (G011.49-01.48 and G0168.06+00.82)  with  large errors  due to weak emission of H$_{2}$$^{34}$S lines. The molecular abundance ratios from \citet {fo23} are also presented.}  
\label{fig:fig3}
\end{figure*}

\section{Results}\label{sec:result}

\subsection{The detected lines} \label{line}

 H$_{2}$S 1$_{10}$-1$_{01}$, H$_{2}$$^{34}$S 1$_{10}$-1$_{01}$, H$_{2}$CS 5$_{14}$-4$_{14}$, HCS$^{+}$ 4-3 and HC$_{3}$N 19-18 were detected in all sources except for  G012.90-00.24, while SiO 4-3 was detected in 46 sources.  C$^{18}$O 1-0  was detected in all sources  (see Figure \ref {fig:fig1} and \ref {figA1}).  H$_{2}$S 1$_{10}$-1$_{01}$ lines are normally stronger  than H$_{2}$CS 5$_{14}$-4$_{14}$, while H$_{2}$CS 5$_{14}$-4$_{14}$ are stronger  than HCS$^{+}$ 4-3. The sources can be divided into three groups based in the line profile of H$_{2}$S 1$_{10}$-1$_{01}$ and   H$_{2}$$^{34}$S 1$_{10}$-1$_{01}$. In 26 sources of the sample both lines can be fitted with a single Gaussian component, such as  G027.36-00.16 (see Figure \ref {fig1:a}), while in 21 sources, H$_{2}$S 1$_{10}$-1$_{01}$ needs from two to four components, and this is probably caused by inflowing gas, such as G031.28+00.06 (see Figure \ref {fig1:b}). There are three sources, G049.48-00.38 (W51M, see Figure \ref {fig1:c}), G043.16+00.01 (W49N) and G000.67-00.03 (SgrB2),  in which there is a possible optically thin velocity component of  H$_{2}$S 1$_{10}$-1$_{01}$ without corresponding H$_{2}$$^{34}$S 1$_{10}$-1$_{01}$ emission. The column densities of H$_{2}$S have been calculated by those of H$_{2}$$^{34}$S for all the sources.

 HC$_{3}$N 19-18 was used to confirm the central velocities and line profile of  S-bearing lines, while  C$^{18}$O was used to get H$_{2}$ column densities to obtain the relative H$_{2}$S,  H$_{2}$CS, and HCS$^{+}$ abundances. SiO was used as the  tracer of shock activities in star formation regions. 
 

 Table \ref {table:source} show the molecular species, velocity-integrated intensities, central velocity, and peak temperature, which are listed in columns 5-8. Velocity integrated  fluxes were obtained from the single component Gaussian fitting, or with ``print area'' for H$_{2}$S, H$_{2}$CS and C$^{18}$O in CLASS if complex line profile was found as explained in Section \ref {me}, which was noted by "a" superscript in  columns 4 of Table \ref {table:source}.    As shown in Table \ref {table:source}, the velocity-integrated intensities and peak temperature of H$_{2}$S 1$_{10}$-1$_{01}$ are higher  than H$_{2}$CS 5$_{14}$-4$_{14}$, while H$_{2}$CS 5$_{14}$-4$_{14}$ are higher  than HCS$^{+}$ 4-3. The S-bearing species studied in this work present similar line width, while SiO 4-3 presents broader line width than other lines in each source.

\subsection{Column densities} \label{col}

The beam averaged  column densities of H$_{2}$S, H$_{2}$$^{34}$S, H$_{2}$CS, HCS$^{+}$, SiO and C$^{18}$O molecules are shown in Table \ref {table:density}, with  the range of  H$_{2}$S from  3.59$\times$10$^{13}$ to 3.50$\times$10$^{15}$cm$^{-2}$, H$_{2}$$^{34}$S from  1.65$\times$10$^{12}$ to 1.83$\times$10$^{14}$cm$^{-2}$
  H$_{2}$CS  from  7.09$\times$10$^{12}$ to 1.25$\times$10$^{15}$cm$^{-2}$ and HCS$^{+}$  from  6.67$\times$10$^{11}$ to 6.04$\times$10$^{13}$cm$^{-2}$, as lowest abundant S-bearing molecules among H$_{2}$S, H$_{2}$CS and HCS$^{+}$.  The resulting column densities  for SiO and C$^{18}$O molecules in these 50 sources were from 3.33$\times$10$^{11}$ to 9.25$\times$10$^{13}$cm$^{-2}$, and from 6.30$\times$10$^{15}$ to 3.76$\times$10$^{16}$cm$^{-2}$, respectively. The column densities of H$_{2}$ are obtained by  C$^{18}$O with  $N_{\rm H_{2}} = 4.37 \times 10^{6} N_{\rm C^{18}O}$ \citep{fre82},  ranging from 1.13$\times$10$^{22}$ to 4.80$\times$10$^{23}$cm$^{-2}$.

\subsection{Relative abundances} \label{abu}

Relative abundances of these molecules, which can be obtained with beam averaged column densities,   are more important than the beam averaged column densities themselves for scientific analyses. 
The abundances of H$_{2}$S,  H$_{2}$CS and  HCS$^{+}$  normalized  by H$_{2}$ were shown in column 4-6 of Table \ref {tableA1}, where H$_{2}$S are obtained from H$_{2}$$^{34}$S multiplied by the $^{32}$S/$^{34}$S.
The relation between the abundances of H$_{2}$S and  HCS$^{+}$  normalized  by H$_{2}$ was presented in Figure \ref {fig2:a}, with Pearson correlation coefficient \footnote{The Pearson correlation coefficient is a measure of the linear relationship between two variables. The closer |$\gamma$| is to 1, the stronger the linear relationship.} of 0.76, slope of 0.97, 
using least square fitting, while  the Pearson correlation coefficients and slope between the abundances of H$_{2}$S and  H$_{2}$CS  normalized  by H$_{2}$ (see Figure \ref {fig2:b}) were 0.77 and 0.98,  respectively. For the abundances of H$_{2}$CS  and HCS$^{+}$ normalized H$_{2}$ (see Figure \ref {fig2:c}), the Pearson  correlation coefficients and slope were  0.94 and 1.00.  Since the slopes and Pearson correlation coefficients among them are are positive numbers  close or equal to 1, H$_{2}$S,  H$_{2}$CS and HCS$^{+}$ molecules are probably very closely chemically related to each other. 

The  relative abundances of H$_{2}$S,  H$_{2}$CS, and HCS$^{+}$  normalized by H$_{2}$ span more than  one order of magnitude in these 50 sources. Table \ref{table:range} showed the  abundance ratio [H$_{2}$S/H$_{2}$CS], [H$_{2}$S/HCS$^{+}$], [H$_{2}$CS/HCS$^{+}$] towards all the sources. The  abundance ratio of [H$_{2}$S/H$_{2}$CS] ranges from 1.32$\pm$0.12 in G000.67-00.03 to 10.2$\pm$0.49 in G109.87+02.11, with median value of  4.4, while it ranges from  14.1$\pm$7.01 in G011.49-01.48 to 129.6$\pm$11.1 in G109.87+02.11 with median value of 63.8 for  [H$_{2}$S/HCS$^{+}$]  and from 7.94$\pm$0.61 in G232.62+00.99 to 23.6$\pm$0.96 in G049.48-00.36 with median value of 13.0 for  [H$_{2}$CS/HCS$^{+}$].

Good correlation can be found between H$_{2}$CS and HCS$^{+}$ whether normalized by H$_{2}$S or H$_{2}$. (see Figure \ref {fig:fig3} and \ref {fig2:c}). The abundance ratios of  [H$_{2}$CS/HCS$^{+}$]    is well within a narrower ranges than that of  [H$_{2}$S/HCS$^+$]  and [H$_{2}$S/H$_2$CS] (see Table \ref {table:range}). G011.49-01.48 and G0168.06+00.82 ,with large errors,   are also marked in Figure  \ref{fig:fig2} and \ref {fig:fig3}, where H$_{2}$$^{34}$S lines showed weak emission.

The relationships between S-bearing molecules and SiO, normalized by H$_{2}$, are shown in  Figure \ref{fig:fig5}, with Pearson correlation coefficients of HCS$^{+}$, H$_{2}$CS, and  H$_{2}$S  with SiO as 0.60, 0.68, 0.57, while slopes are 0.73, 0.75, 0.72, respectively. Even though the correlations are not as good as that of in between the S-bearing molecules themselves (see Figure \ref {fig:fig2}), the abundances of HCS$^{+}$, H$_{2}$CS, and  H$_{2}$S  increase with the increment of SiO  abundance in these sources.  Shock chemistry might be needed for further modeling.



\subsection{Line profiles} \label{line}
In most cases, the lines are very well fitted with a single Gaussian (see Figure \ref {fig:fig1} and \ref {figA1}). This is especially apparent in H$_{2}$$^{34}$S,  H$_{2}$CS, HCS$^{+}$ and HC$_{3}$N, except for H$_{2}$S and SiO. The values of full width at half maxima (FWHM) are similar to each other for H$_{2}$$^{34}$S,  H$_{2}$CS, HCS$^{+}$ and HC$_{3}$N lines in each source,   which are between 1 km s$^{-1}$  and 10 km s$^{-1}$  in all sources (see Table \ref {table:fwhm}), except for G000.67-00.03 (Sgr B2) and G043.16+000.01 (W49N). Figure \ref{b} shows comparison among the  FWHMs of the observed lines, for  which the correlation is perfect between HCS$^{+}$ and H$_{2}$CS with Pearson correlation coefficient of 0.98, good between H$_{2}$CS and H$_{2}$$^{34}$S (0.94), HCS$^{+}$ and H$_{2}$$^{34}$S (0.95), HCS$^{+}$ and HC$_{3}$N (0.92).  Such results indicate that  these observed  H$_{2}$$^{34}$S,  H$_{2}$CS, HCS$^{+}$ and HC$_{3}$N  transitions trace the same gas.

\begin{figure*}
 \centering 
\subfigure[]{ \label{fig5:a} 
 \includegraphics[width=0.65\columnwidth]{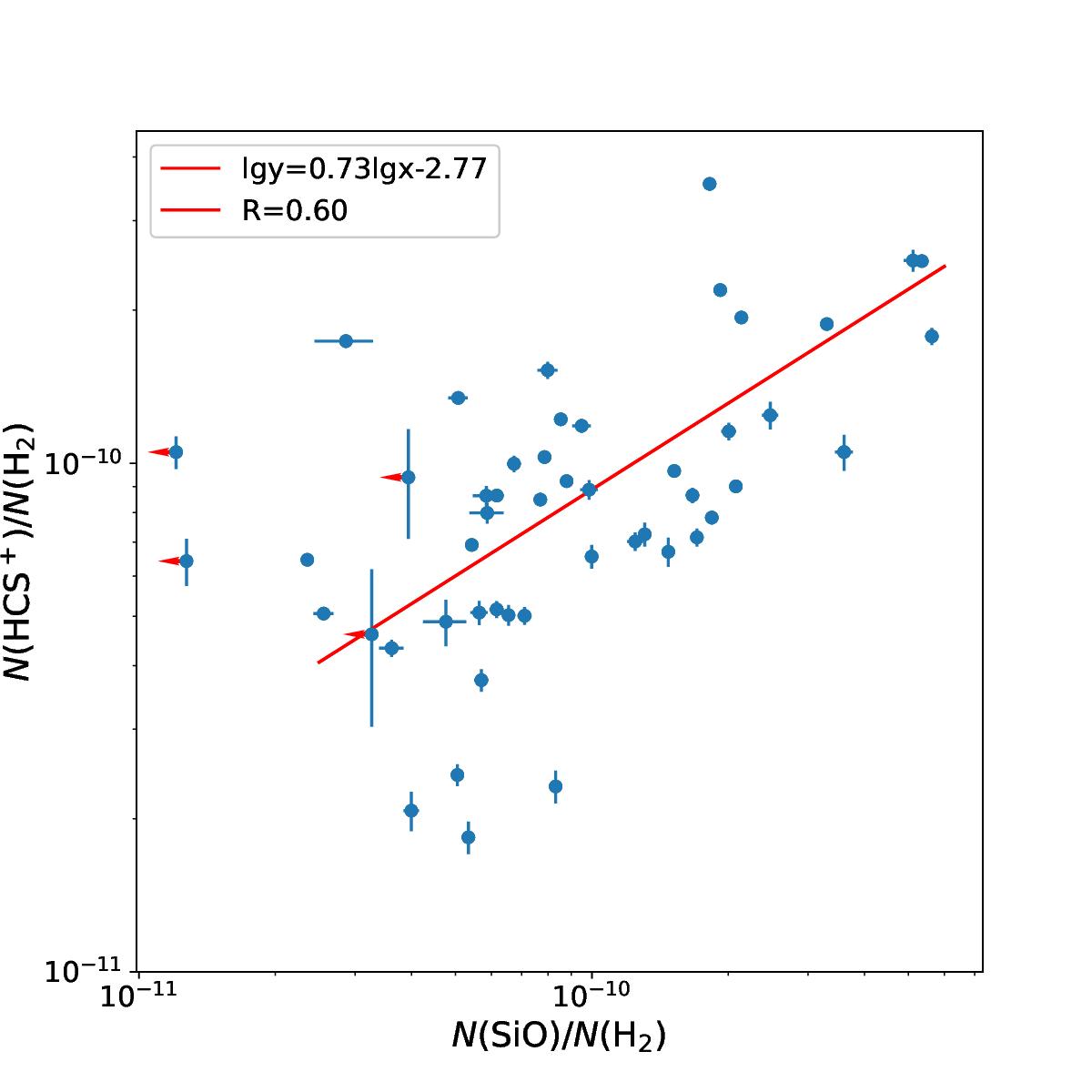} 
} 
\subfigure[]{ \label{fig5:b} 
\includegraphics[width=0.65\columnwidth]{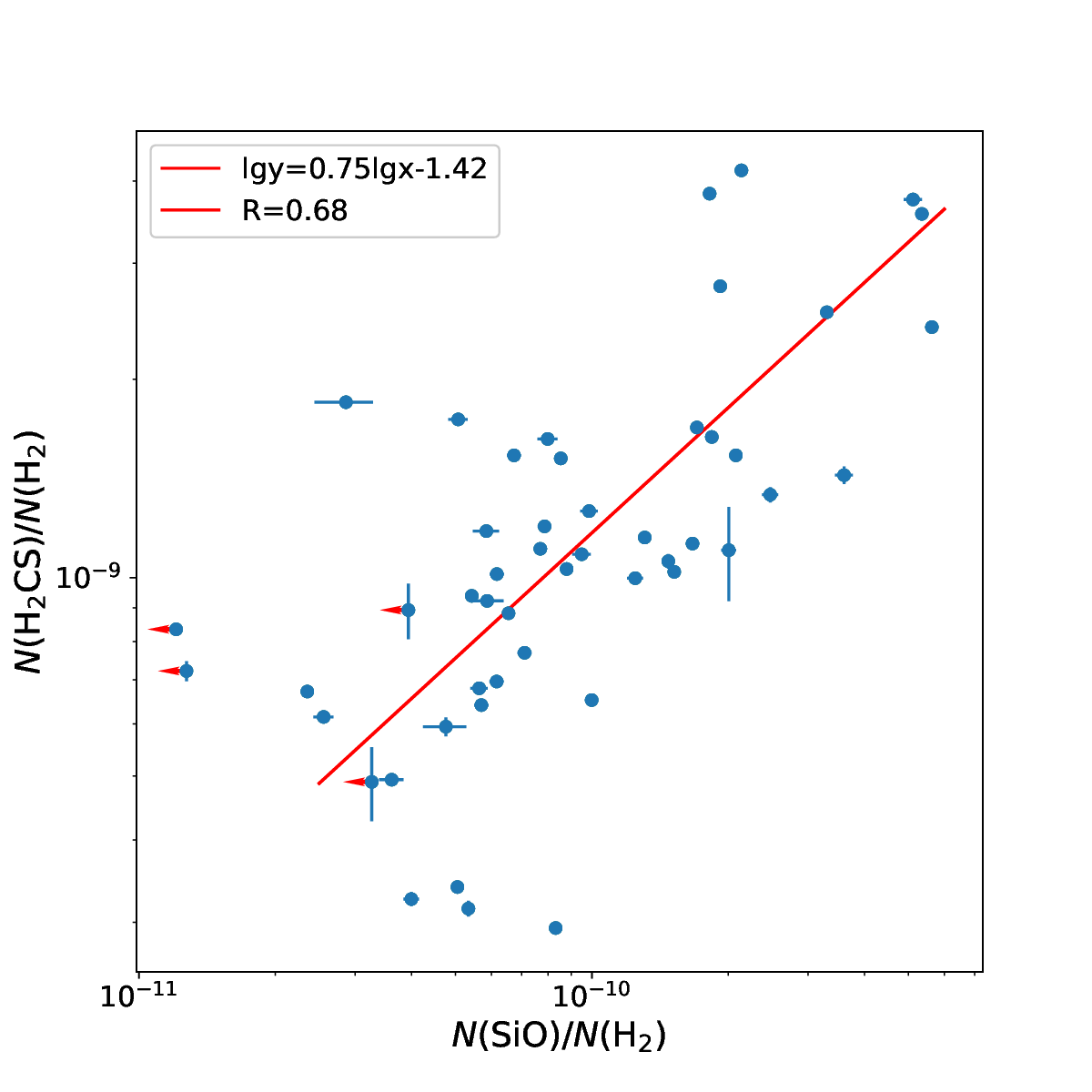}
} 
\subfigure[]{ \label{fig5:c} 
\includegraphics[width=0.65\columnwidth]{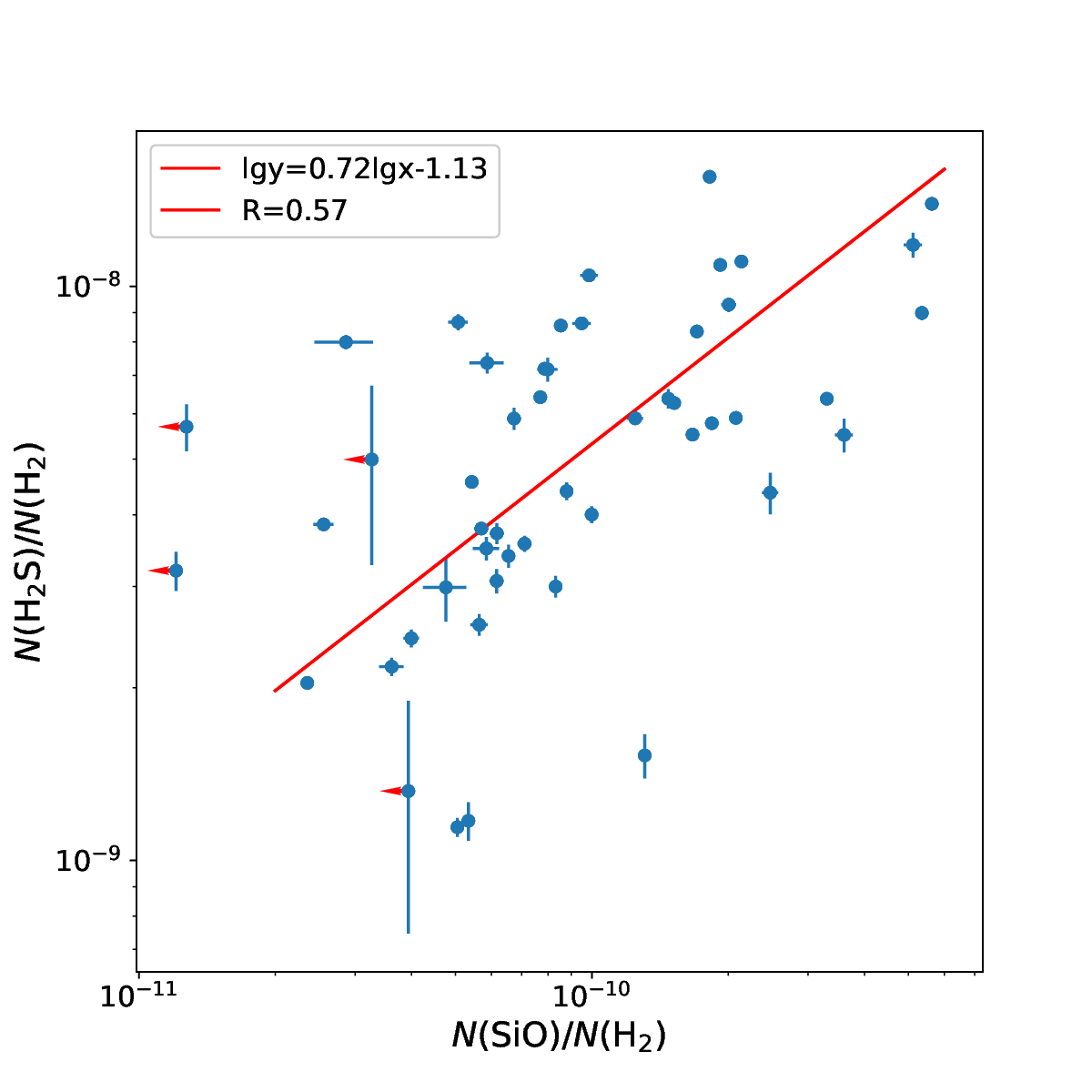}
} 
\caption{The relationship between the three molecules and SiO, which use H$_{2}$ to normalize. (a) shows the relationship between HCS$^{+}$ and SiO abundances. And (b) shows relationship between the H$_{2}$CS and SiO abundances, while (c) shows the relationship between H$_{2}$S and SiO abundances. The fitting lines are marked in red, using the least square method. The four points marked with red arrows are realistic upper limits of SiO, which are no involvement in fitting.}  
\label{fig:fig5}
\end{figure*}

\begin{figure*}
\centering
\subfigure{
\includegraphics[width=0.5\columnwidth]{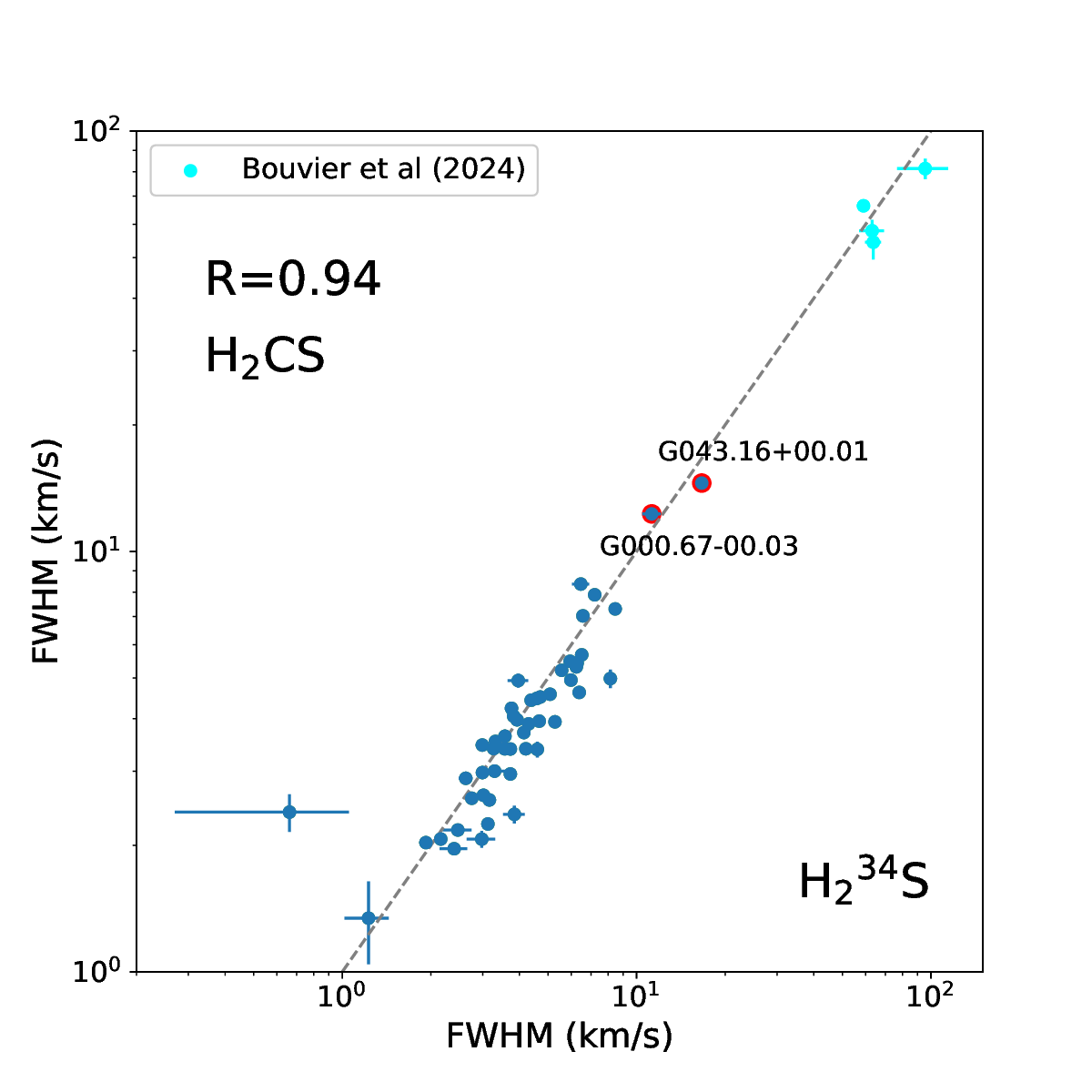} \includegraphics[width=0.5\columnwidth]{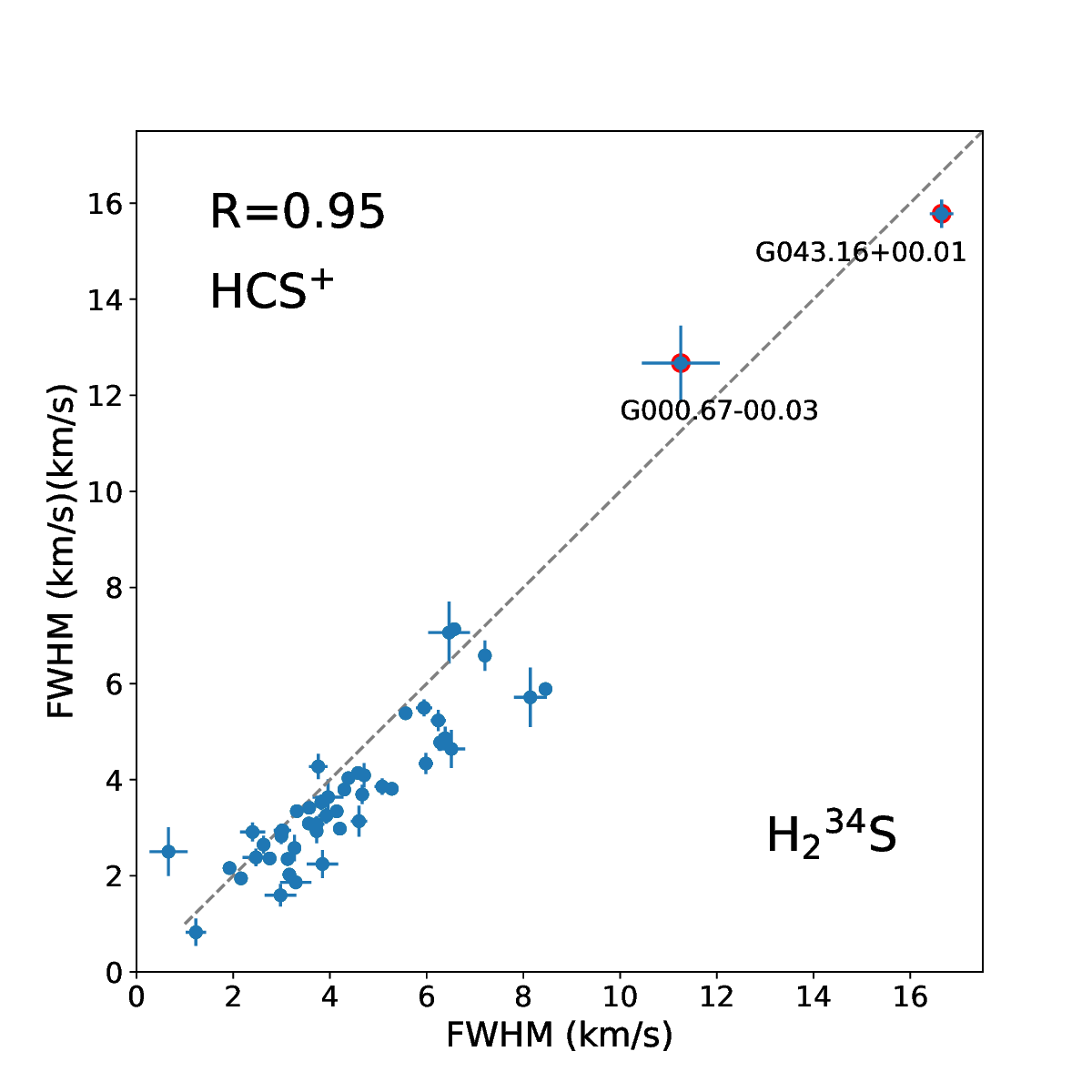} \includegraphics[width=0.5\columnwidth]{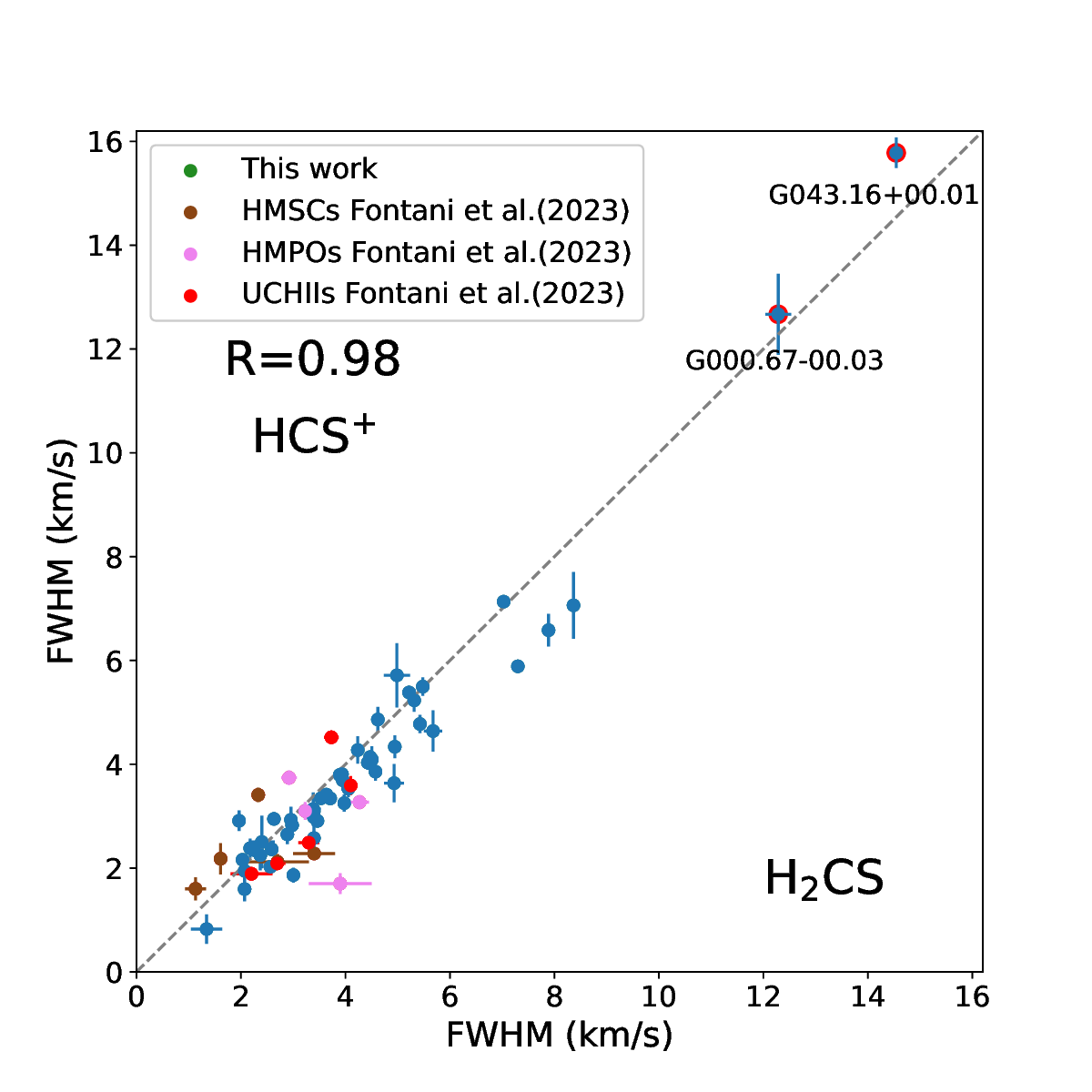} \includegraphics[width=0.5\columnwidth]{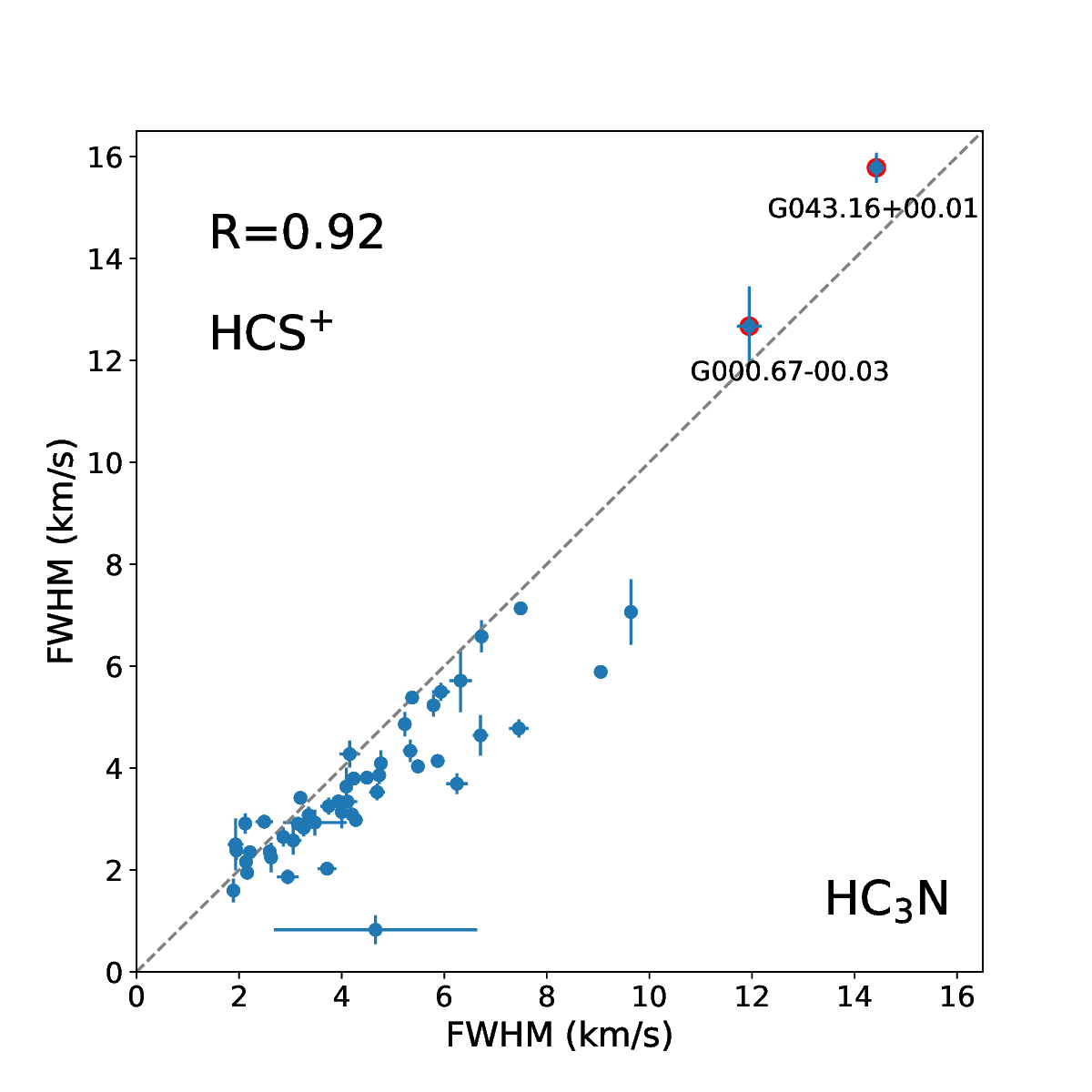} 
} 
\caption{Comparison among the lines FWHMs of the observed species. In all plots, the two blue points with red circles that are far away are G000.67-00.03 (Srg B2) and G043.16+000.01 (W49N). The number in the upper left corner of each panel is the Pearson $\rho$ correlation coefficient. The gray line is y=x. The ratios from \citet {bou24} and  \citet {fo23}  are also presented.}  \label{b} 
\end{figure*}

\begin{figure*}
 \centering 
\includegraphics[width=0.8\textwidth]{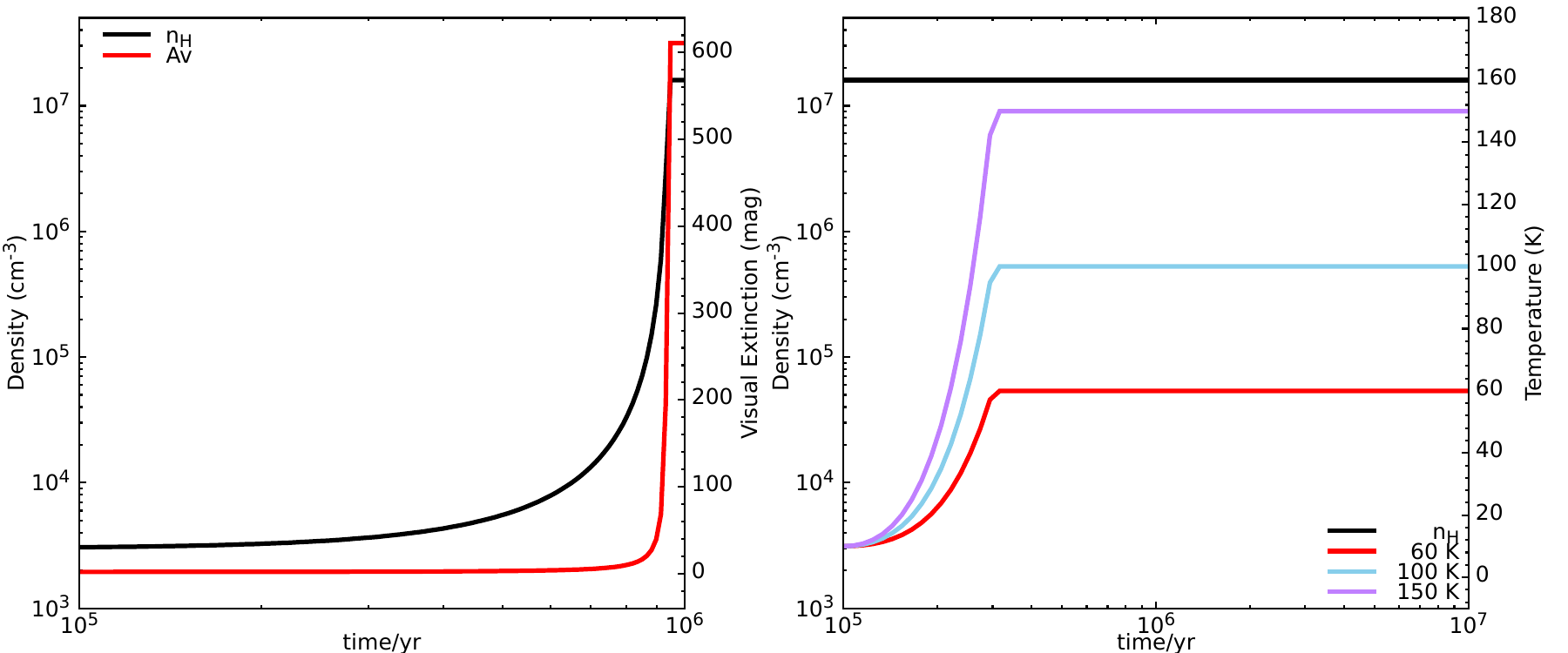}

\caption{The density, temperature, and A$_{V}$ profiles as functions of time for collapse and warm-up stages in hot core models. The left panel represents the collapse stage, during the stage, the gas density gradually increases from 3$\times$10$^{3}$ cm$^{-3}$ to 1.6$\times$10$^{7}$ cm$^{-3}$ over a period of 1$\times$10$^{6}$ years, and the temperature remains at a constant value of 10 K. An increase in gas density leads to an increase in visual extinction. The right panel represents the warm-up stage, during the stage, the gas density and visual extinction are 1.6 $\times$10$^{7}$ cm$^{-3}$ and 6.109$\times$10$^{2}$ mag, respectively, T$_{max}$ adopt three values, 60, 100 and 150K.} 
\label{para}

\end{figure*}

\begin{figure*}
 \centering 
\subfigure[]{ \label{fig4:b} 
\includegraphics[width=0.65\columnwidth]{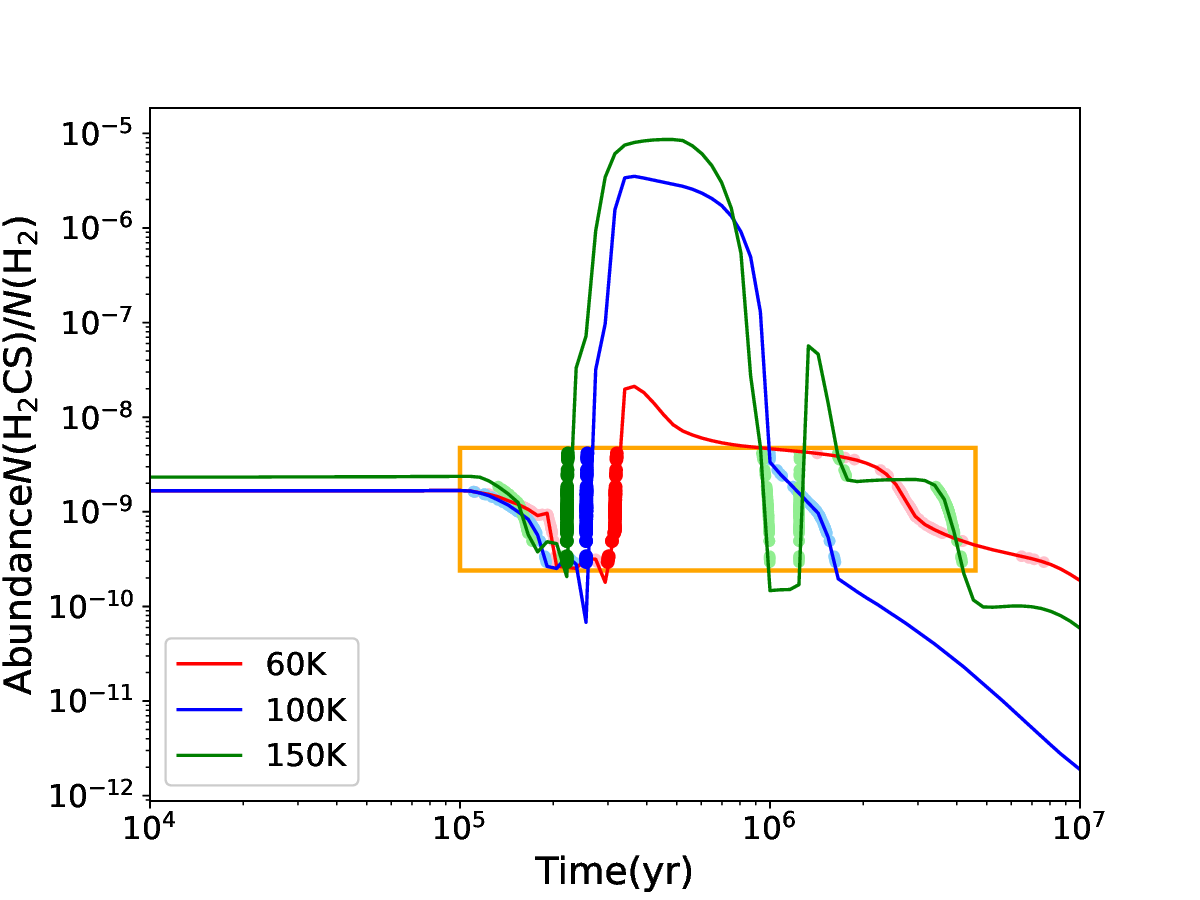}
} 
\subfigure[]{ \label{fig4:c} 
\includegraphics[width=0.65\columnwidth]{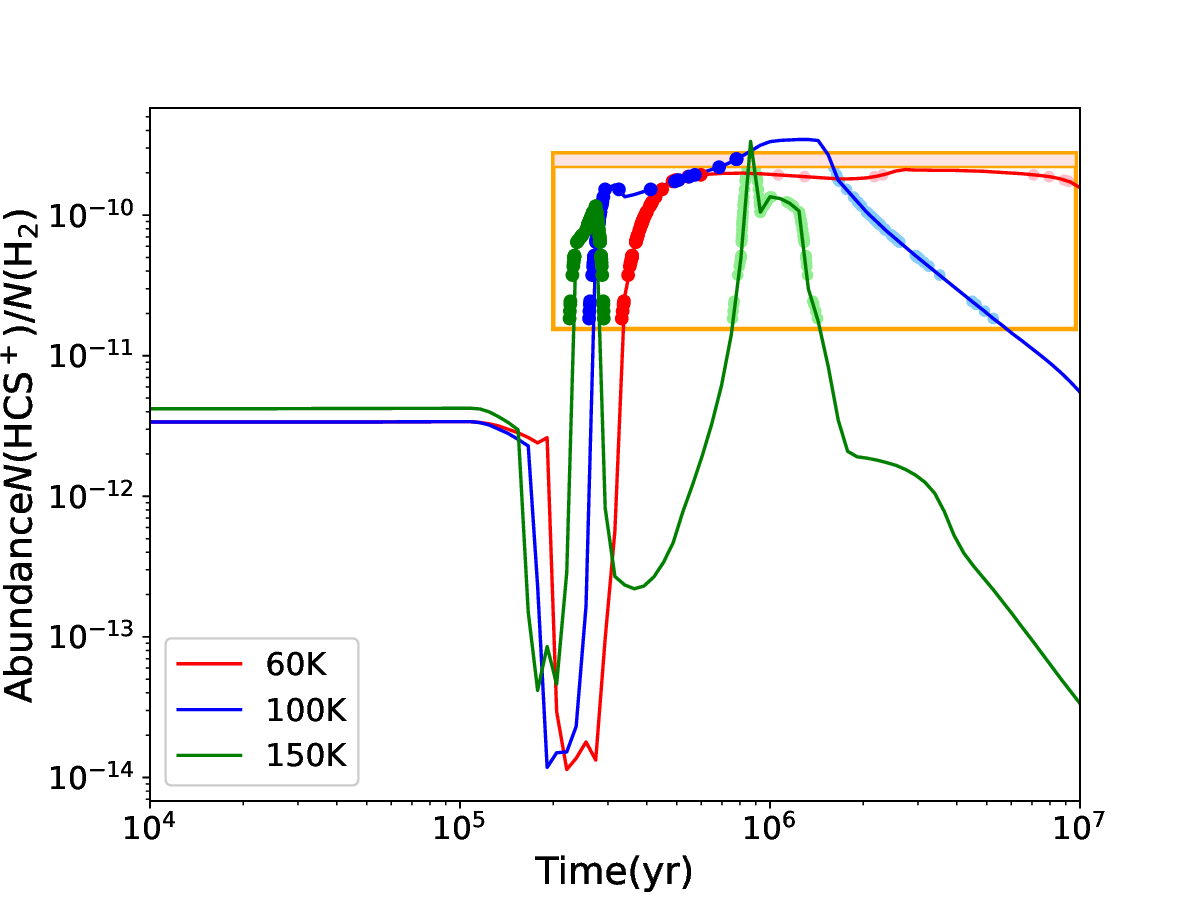}
} 
\subfigure[]{ \label{fig4:a} 
 \includegraphics[width=0.65\columnwidth]{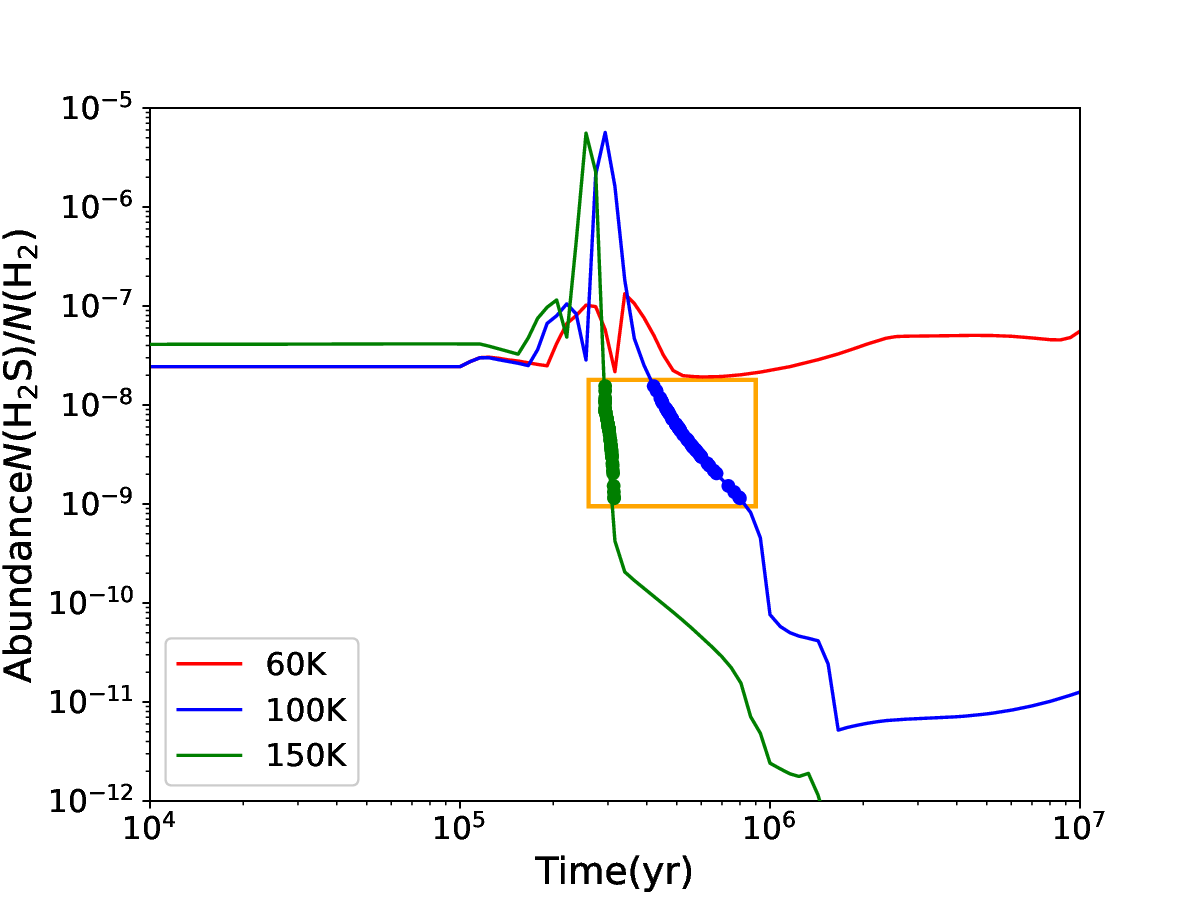} 
} 
\caption{Chemical model including includes the gas, the dust grain surface and the icy mantle. 
The solid lines correspond to the gas phase abundances versus time during warm-up stages in chemical models with gas maximum temperatures at 60, 100, and 150K, and the points are the observed datas. The abundances of  H$_{2}$CS,  HCS$^{+}$ and H$_{2}$S, relative to H$_2$, are presented in (a), (b) and (c), respectively.  Since the samples are in the same evolutionary stage, two possible time are marked with filled and transparent circles from model. Note that there are  no intersections for H$_{2}$S at 60K in model.}
\label{fig:fig4}
\end{figure*}

\section{Discussion}\label{sec:dis}

\subsection{Possible chemical connections between {\rm H$_2$S, H$_2$CS}, and {\rm HCS$^+$} in hot cores?} \label{rela}

S-bearing species are good tracers of hot cores because they are particularly sensitive to physical and chemical changes in dense and hot regions  \citep{hat98}. H$_{2}$S is the main molecular form of S-bearing molecules  in  molecular clouds, which may be produced on the surface of grains, and the rapid reactions of H$_2$S molecules can  drive the production of other S-bearing molecules  \citep{wak12}. The measured abundances of  H$_2$S:H$_2$CS:HCS$^+$ in these 50 sources  showed that H$_2$S is the most abundant molecule with respect to  H$_2$S, H$_2$CS and HCS$^+$, what is found for H$_2$O, more abundant than H$_2$CO, and HCO$^+$  \citep{her73}.



The abundances of HCS$^+$ and H$_{2}$CS presented even tighter correlation than that of these two molecules and H$_2$S with correlation coefficient of 0.94 (see Figure \ref{fig2:c}). 
This is likely due to diﬀerent chemical formation processes of H$_2$S compared to HCS$^+$ and H$_2$CS in chemical models. H$_2$S is mainly formed on the grain surface \citep{wak12}, while the other two molecules are primarily formed in the gas phase \citep {vidal18, gro14}. Additionally, the elemental carbon also affects the formation of H$_2$CS and HCS$^+$, which can reﬂect to the better correlation of  HCS$^+$ v.s. H$_2$CS than that of HCS$^+$ v.s. H$_2$S and H$_2$CS v.s H$_2$S in observation.
Not only the correlation coefficients,   but also the slopes of fitting results  can be used to justify the relation between the abundances two molecules.  We find a tight linear relationship between HCS$^+$ and H$_2$CS with slope of 1.00, while that of the other two are 0.98 for H$_{2}$CS v.s. H$_2$S and 0.97 for HCS$^+$ v.s. H$_2$S.

Our sample  included 51 massive star forming regions at late stage, while  the sample in  \citet {fo23} (see Figure \ref{fig:fig3}) included 15 sources at different stages. The HCS$^{+}$/H$_{2}$S v.s. H$_{2}$CS/H$_{2}$S trend found in this work is also followed by the sources observed in \citet {fo23}.
The line widths and excitation temperatures were obtained to  explain that HCS$^{+}$  trace preferentially quiescent and likely extended material. Our results showed that  line widths of H$_{2}$$^{34}$S,  H$_{2}$CS, HCS$^{+}$ and HC$_{3}$N lines in each source were almost the same  (see Figure \ref{b}), which indicate that they may be from the same gas.

\subsection{Chemical model   including   gas,  dust grain surface and  icy mantle phases, for { \rm H$_2$S, H$_2$CS}, and {\rm HCS$^+$} molecules } \label{che}

In order to understand the observed abundances of H$_{2}$S, H$_{2}$CS and HCS$^{+}$ in hot cores, we utilize the three-phase NAUTILUS chemical code  \citep {rua16},  including  gas,  dust grain surface and icy mantle. This is combined with a similar physical model in  \citet {zha23}, where a collapse stage is followed by a warm-up stage. During the collapse stage, the gas density is initially set at 3$\times$10$^{3}$ cm$^{-3}$ and gradually increases to the final density 1.6$\times$10$^{7}$ cm$^{-3}$ over a period of 1$\times$10$^{6}$ years. The gas temperatures are fixed at 10 K during this stage. The density gradually increases over time, resulting in an increase in extinction, as shown in the left panel of Fig \ref {para}. Once the peak density is achieved, the density is fixed and the gas and dust temperatures are raised from $\sim$10 K to their peak temperatures in 2$\times$10$^{5}$ years during the warm-up stages, as shown in the right panel of Fig \ref {para}. The main physical parameters of hot core models are summarized in Table \ref {table:model}. For our simulations, the initial abundances are those of Table 1 of  \citet {vidal18}. The chemical network is based on the research work of  \citet {vidal17}.  

In warm-up stage, H$_{2}$S is mainly formed through the following hydrogenation reactions on grain surfaces:
\begin{equation}
{\rm S + H  \rightarrow HS + H  \rightarrow H_{2}S}.
\end{equation}

\noindent Some H$_{2}$S is already released back into the gas phase through reactive desorption before temperature rising. When the temperature reaches high values after 2$\times$10$^{5}$ yr, it can return to the gas phase through thermal process, which plays a dominant role in the production of gaseous H$_{2}$S. Initially, H$_{2}$S is mainly destroyed by H$_{3}$$^{+}$ via ion-neutral reactions in all cases. However, after 2$\times$10$^{5}$ yr, 7$\times$10$^{5}$ yr, and 3$\times$10$^{5}$ yr, it is mainly destructed by atomic carbon for a T$_{max}$ of 60 and 100 K, and by atomic hydrogen for T$_{max}$ of 150 K in the gas phase, respectively. The related reactions are as follows: 

\begin{equation}
\begin{split}
{\rm H_{2}S + H_{3}^{+}  \rightarrow H_{2} + H_{3}S^{+} ; \quad C + H_{2}S  \rightarrow HCS + H ;}  \\  
{\rm  H + H_{2}S  \rightarrow H_{2} + HS.}
\end{split}
\end{equation}

\noindent H$_{2}$CS is primarily formed  in gas phase through the following neutral-neutral reaction:
\begin{equation}
{\rm S + CH_{3}  \rightarrow H_{2}CS + H}.
\end{equation}
\noindent and it follows a similar destruction pathway to H$_{2}$S, as both are initially broken down through the ion H$_{3}$$^{+}$. After 5$\times$10$^{5}$ yr, 7$\times$10$^{5}$ yr and  9$\times$10$^{5}$ yr, respectively, in the gas phase, it is mainly destroyed by atomic carbon for T$_{max}$ of 60, 100, and 150 K, respectively. The reactions proceed as follows:

\begin{equation}
{ \rm H_{2}CS + H_{3}^{+}  \rightarrow H_{2} + H_{3}CS^{+} ; \quad  H_{2}CS + C \rightarrow HC_{2}S + H}.
\end{equation}

\noindent Initially, HCS$^{+}$ is mainly formed through the reaction of CS with the H$_{3}$$^{+}$ ion in all cases. However, after 3$\times$10$^{5}$ yr and 7$\times$10$^{5}$ yr, it is primarily produced via the reactions CS + HCO$^{+}$ and CS + HCNH$^{+}$ for T$_{max}$ of 60 and 150 K, respectively. The related reactions are as follows:


\begin{equation}
\begin{aligned}
\rm CS + H_{3}^{+}  \rightarrow HCS^{+} + H_{2}  ; \quad CS + HCO^{+}   \rightarrow CO + HCS^{+} ; \\  
 \rm CS + HCNH^{+}   \rightarrow HCN + HCS^{+}.
\end{aligned}
\end{equation}

\noindent It is mainly destroyed in gas phase by the following electron recombination:

\begin{equation}
{\rm HCS^{+} + e^{-}  \rightarrow  S + CH}.
\end{equation}

Both H$_2$S and H$_2$CS  can be destroyed  by the reactions with H$_3$$^+$ (see Eq. (6) and (8)), while it is the main formation path of HCS$^+$ by  reaction with H$_3$$^+$ (see Eq. (9)).  The relation of H$_2$S, H$_2$CS and HCS$^+$ with  H$_3$$^+$ can cause close connection among them. 


If we compare with the similar O-bearing species, H$_{2}$CO could be produced from H$_{2}$O+CO, and is the second most abundant product after  CO$_{2}$, at different concentrations of H$_{2}$O:CO mixtures  \citep {bar22}. There are other processes to  form  H$_{2}$CO from  H$_{2}$O, such as H$_{2}$O+C $\to$ H$_{2}$CO at temperatures associated with ISM and the cooler regions of the environment around young stars  \citep{pot21}. It shows that H$_{2}$CO is formed in a different way with H$_{2}$CS. The formation of HCO$^{+}$ under different A$_{v}$ is different. From A$_{v}$ $\sim$ 0.4 to 3, the dominant reaction is CO$^{+}$ + H$_{2}$, while above 3, the dominant reaction is H$_{3}$$^{+}$ + CO in molecular clouds \citep {pan23}. HCO$^{+}$ can be destroyed by C$_{3}$H$_{2}$ \citep {na24}. The formation paths of HCO$^{+}$ and HCS$^{+}$ seem to be the same in particular cases.


 
Considering processes including what have been described before, the results of relative   abundances of H$_{2}$S, H$_{2}$CS and HCS$^{+}$ species in gas phase  varying with the time   for warm-up stages from the model are presented in Figure \ref {fig:fig4}, with  three  different maximum temperatures (T$_{max}$) as 60K, 100 K and  150K.   The   measured abundances are compared with the abundances  from the model, which can have  same value at different time (see Figure  \ref {fig:fig4}). Since the samples are in the same evolutionary stage, the more reasonable intersections are marked with filled circles. Transparent circles show another possible time.
  We considered the calculated abundance consistent with the observed value if the  calculated abundance differs from the observed value within one order of magnitude  value.

For T$_{max}$=60K (see Figure \ref {fig:fig4}, red lines), the abundance of H$_{2}$S and H$_{2}$CS varies much less than that of HCS$^{+}$. Besides, there is one possible time (3$\times10^5$ yr)  for each source that match  the measured [H$_{2}$CS/H$_{2}$] and [HCS$^{+}$/H$_{2}$] (red line and filled circles) abundances (see  Figure \ref{fig4:b}, \ref {fig4:c}), while no intersection for [H$_{2}$S/H$_{2}$] (see Figure \ref{fig4:a}). However, the discrepancy is less one order of magnitude. Comparing to the observational results, the over prediction of H$_{2}$S that we find could be a result of the absence of effective gas-phase destruction reactions in the H$_{2}$S chemistry within our model. [HCS$^{+}$/H$_{2}$] abundances in all sources range from 1.84$\times$10$^{-11}$ to 3.54$\times$10$^{-10}$, and four of them are above the model results  from 2.19$\times$10$^{-10}$ to 3.54$\times$10$^{-10}$ (see  Figure \ref{fig4:c}, orange rectangle box with fill), while all sources with measured  [H$_{2}$CS/H$_{2}$]  abundances, from 2.94$\times$10$^{-10}$ to 4.15$\times$10$^{-9}$, are matched the model results.  

The abundances of  H$_{2}$CS and HCS$^{+}$  changes quickly than  H$_{2}$S in 1$\times$10$^{5}$ to 1$\times$10$^{7}$ yr, with T$_{max}$=100K (see Figure \ref {fig:fig4}, blue lines).  There is not a single time for comparisons between the observed value and the model.  The solutions are  totally different for different molecules  (see Figure  \ref {fig:fig4}), with about  2$\times10^5$ yr for [H$_{2}$CS/H$_{2}$], 3$\times10^5$ yr for  [HCS$^{+}$/H$_{2}$] and 5$\times10^5$ yr for [H$_{2}$S/H$_{2}$].

The possible time with T$_{max}$=150K  for each  source to match  the measured abundances  of [H$_{2}$S/H$_{2}$]  is concentrated around 3$\times$10$^5$ yr (see Figure \ref{fig4:a}, green  lines). For the two to four possible solutions of time in each source for [H$_{2}$CS/H$_{2}$]  and  [HCS$^+$/H$_{2}$], it is more likely to be the younger one for most of the sources, with one time for each to explain  the three different abundances in each source,  around 3$\times$10$^5$ yr. 

In general, the models can reproduce the observed abundances at around 2-3$\times$10$^{5}$ yr for almost all of the sources. However, there are  small discrepancy at lower T$_{max}$ (60K) models for H$_{2}$S. H$_{2}$CS and HCS$^{+}$ can be efficiently formed in gas-phase, while H$_{2}$S already chemically desorbs during its formation on the grains before being thermally desorbed. The abundance of H$_{2}$S cannot sharply increase like at T$_{max}$ of 100K and 150K after the temperature reaches T$_{max}$ of 60K because it does not reach the desorption temperature of H$_{2}$S (around 70K). In this condition, the chemical desorption is primary origin for gaseous H$_{2}$S. Since gas-phase destruction reactions for H$_{2}$S are not as effective at relatively low temperature of 60K compared to T$_{max}$ of 100K and 150K, the abundance of H$_{2}$S is over predicted in our model.


\subsection{Shock activities} \label{sho}

SiO was considered to be an excellent tracer of shock activities in star formation regions, because SiO abundance in molecular outflows was increased by at least three orders of magnitude compared to those in quiescent gas  \citep {97sch}. Lots of  elemental Si were probably to be found in the form of stable materials in dust, like silicates, which were difficult to release Si atoms. Shock activities were needed to destroy silicates to release Si atoms from the grains into gas phase  \citep {92ma}, while S-bearing molecules can be enhanced with moderate shocks  \citep{wak04}.

SiO 4-3 was detected in 46 of 51 sources in our sample (see green lines  in Figure \ref {fig:fig1} (d), (e) and (f)), indicating shock activities there.  The abundances of HCS$^{+}$, H$_{2}$CS, and  H$_{2}$S  increase with the increment of SiO  abundance in these sources (see Figure \ref {fig:fig5}), which implies that shock chemistry may be important for  HCS$^{+}$, H$_{2}$CS, and  H$_{2}$S. In addition, the line width of SiO is larger than those of the S-bearing species. This  may indicates that the SiO is actually tracing a different physical component with a different velocity broadening.  Indeed, SiO likely traces shocked regions, which implies that the S-bearing species are either tracing weaker shocks not traced by SiO or are actually tracing the hot core regions.

\section{Summary and conclusions}\label{sec:conclusion}
We have observed H$_{2}$S 1$_{10}$-1$_{01}$, H$_{2}$$^{34}$S 1$_{10}$-1$_{01}$, H$_{2}$CS 5$_{14}$-4$_{14}$, HCS$^{+}$ 4-3, HC$_{3}$N 19-18,  SiO 4-3 and C$^{18}$O 1-0 towards a sample of 51 massive star forming regions with the IRAM 30-m millimeter telescope. We have obtained beam average column densities, as well as their abundances with respect to H$_{2}$, derived from C$^{18}$O.

All lines can be  well fitted with a single Gaussian, except for H$_{2}$S and SiO. The values of FWHMs are between 1 km s$^{-1}$  and 10 km s$^{-1}$  in all sources, except for G000.67-00.03 (Srg B2) and G043.16+000.01 (W49N).  Comparising among the lines FWHMs of the observed species,  we find that they are positively correlated with each other, as well as  they have  similar FWHM ranges, which indicate that they may trace similar regions.

The abundance ratios of  [H$_{2}$S/H$_{2}$], [H$_{2}$CS/H$_{2}$] and [HCS$^{+}$/H$_{2}$]  range by more than one magnitude with high correlation coefficients and positive slopes  between them, with correlation coefficient of 0.94, slope of 1.00 for  [H$_{2}$CS/H$_{2}$] and [HCS$^{+}$/H$_{2}$], while the correlation coefficient and slope between  [H$_{2}$S/H$_{2}$] and [HCS$^{+}$/H$_{2}$] are 0.76 and 0.97, respectively. For  [H$_{2}$CS/H$_{2}$] and [H$_2$S/H$_{2}$], the correlation coefficients and slope are 0.77 and 0.98.
   The close relations  indicate that these S-bearing molecules may have chemical connections with each others, with HCS$^+$ and H$_{2}$CS the most correlated ones, based on the relations between the relative abundances of these molecules in the 50 sources. 

The three-phase NAUTILUS chemical code was utilized to simulate the relationship of three S-bearing molecules. Comparing the observed abundance with model results, there are one possible time (2-3$\times$10$^{5}$ yr) for each  source in the model to match  the measured abundances of H$_{2}$S, H$_{2}$CS and HCS$^{+}$.  The abundances of HCS$^{+}$, H$_{2}$CS, and  H$_{2}$S  increase with the increment of SiO  abundance in these sources, which implies that shock chemistry may be important for  them.  Shock activity need to be considered in further modeling H$_{2}$S, H$_{2}$CS and HCS$^{+}$ in hot cores.

\begin{acknowledgements}
This work is supported by National Key R\&D Program of China (2023YFA1608204) and   the National Natural Science Foundation of China grant 12173067. This study is based on observations
carried out under project number 012-16, 023-17 and 005-20  with the IRAM 30m telescope. IRAM is
supported by INSU/CNRS (France), MPG (Germany) and IGN (Spain). 
\end{acknowledgements}

\clearpage
\onecolumn

\begin{table}
\centering
\begin{threeparttable}

\caption{Information of the sources and  detected lines.} \label{table:source}

\begin{tabular}{cccccccc}

\hline
Source name      & RA(J2000)     & DEC(J2000)  & D$_{GC}$& Line       & $\int$T$_{mb}$dv   &   v$_{LSR}$        & T$_{peak}$  \\
            Alias                          & $(hh:mm:ss)$      & $(dd:mm:ss)$      & $(kpc)$ &                       &($K km$ $s^{ -1}$)                                      & ($km$ $s^{ -1}$)  & $(K)$     \\
\hline

G121.29+00.65   & 00:36:47.35  & 63:29:02.2  & 8.8 & HC$_{3}$N (19-18)                                              & 1.92$\pm$0.03    & -17.4$\pm$0.1 & 0.69       \\
L1287          &              &  &            & $^{a}$H$_{2}$S (1$_{10}$-1$_{01}$)                             & 9.08$\pm$0.04    & -17.7$\pm$0.1 & 2.03        \\
               &              &       &       & H$_{2}$$^{34}$S (1$_{10}$-1$_{01}$) & 0.69$\pm$0.03    & -17.5$\pm$0.1 & 0.24       \\
               &              &        &      & $^{a}$H$_{2}$CS (5$_{14}$-4$_{14}$)                                  & 2.27$\pm$0.02    & -17.4$\pm$0.1 & 0.81       \\
               &              &       &       & HCS$^{+}$ (4-3)                                                & 0.93$\pm$0.02    & -17.5$\pm$0.1 & 0.37       \\
               &              &       &       & $^{a}$C$^{18}$O (1-0)                                                & 7.06$\pm$0.02    & -17.6$\pm$0.1 & 2.51       \\
               &              &        &      & SiO (4-3)                                                      & 1.68$\pm$0.06    & -17.1$\pm$0.1 & 0.42       \\
G123.06-06.30   & 00:52:24.70  & 56:33:50.5  & 10.1 & HC$_{3}$N (19-18)                                              & 2.77$\pm$0.03    & -30.8$\pm$0.1 & 0.81       \\
NGC281         &              &      &        & H$_{2}$S (1$_{10}$-1$_{01}$)                                   & 10.40$\pm$0.03   & -31.0$\pm$0.1 & 2.07       \\
               &              &           &   & H$_{2}$$^{34}$S (1$_{10}$-1$_{01}$) & 0.95$\pm$0.02    & -30.5$\pm$0.1 & 0.26       \\
               &              &       &       & H$_{2}$CS (5$_{14}$-4$_{14}$)                                  & 4.70$\pm$0.02    & -30.4$\pm$0.1 & 1.24       \\
               &              &        &      & HCS$^{+}$ (4-3)                                                & 1.68$\pm$0.03    & -30.6$\pm$0.1 & 0.46       \\
               &              &         &     & $^{a}$C$^{18}$O (1-0)                                                & 5.87$\pm$0.03    & -30.0$\pm$0.1 & 1.62       \\
               &              &          &    & SiO (4-3)                                                      & 7.50$\pm$0.08    & -31.0$\pm$0.1 & 0.42       \\
\hline

\end{tabular}
\begin{tablenotes}
\footnotesize
\item  Notes. More information is presented in Zenodo. \url{https://doi.org/10.5281/zenodo.13937627}\end{tablenotes}
\end{threeparttable}
\end{table}

\begin{table}
\centering
\begin{threeparttable}

\caption{Spectroscopic information from CDMS.} \label{table:spectro}

\begin{tabular}{ccccccc}

\hline

Line$^{(a)}$   &   Frequency(MHz)   &    A$_{ul}$(s$^{-1}$)    &  Q(T$_{ex}$=18.75K)    &   g$_{u}$    &  E$_{u}$(cm$^{-1}$)    & Ref.$^{(b)}$ \\
\hline

HC$_{3}$N 19-18 & 172849.300 & 4.08$\times$10$^{-4}$   & 86.2 & 39 & 27.88  & \citep {ka02}\\
H$_{2}$$^{34}$S 1$_{10}$-1$_{01}$  & 167910.516 & 2.64$\times$10$^{-5}$   &  8.72  &  9  & 27.83 &\citep {min91}  \\
H$_{2}$CS 5$_{14}$-4$_{14}$ & 169114.160 & 6.68$\times$10$^{-5}$  &   91.2 &  33  &  37.54   &\citep {th72} \\
HCS$^{+}$ 4-3 & 170691.603 & 9.86$\times$10$^{-5}$  & 18.7 & 9 & 20.48   & \citep {gu81}\\
SiO 4-3 &  173688.274 & 2.61$\times$10$^{-4}$ & 18.3 &  9  &   20.84 & \citep{sc82}  \\
C$^{18}$O 1-0  &  109782.173  & 6.27$\times$10$^{-8}$ &  7.46 & 3 & 5.27   & \citep {ul76} \\

\hline

\end{tabular}
\begin{tablenotes}
\footnotesize
\item  Notes. $^{a}$ All parameters are taken from the CDMS catalogue, $^{b}$  Reference paper
where the observations are presented.
\end{tablenotes}
\end{threeparttable}
\end{table}

\begin{table}
\begin{threeparttable}
\caption{Physical parameters of hot core models.} \label{table:model}

\begin{tabular}{cccccc}

\hline

Source(Stage)   &   $n_{H}$(cm$^{-3}$)   &    T(K)   &  A$_{V}$(mag)    &   $\zeta$(s$^{-1}$)    &UV factor(Habing)    \\
\hline

The free fall collapse$^{a,b}$  &    3$\times$10$^{3}$ $\rightarrow$ 1.6$\times$10$^{7}$   &  10    &   2 $\rightarrow$ 6.109$\times$10$^{2}$  &  1.3$\times$10$^{-17}$   &  1  \\
The warm-up$^{b,c}$  &   1.6$\times$10$^{7}$  &  10 $\rightarrow$ 60, 100, 150   &   6.109$\times$10$^{2}$  &  1.3$\times$10$^{-17}$   &  1  \\
     
\hline

\end{tabular}

\begin{tablenotes}
\footnotesize
\item  Notes. $^{a}$  \citet{ga06}, $^{b}$  \citet {bon19}, $^{c}$  \citet {cou18} 
\end{tablenotes}
\end{threeparttable}

\end{table}

\onecolumn
\begin{appendix}
\section{Spectra for individual sources} 
\begin{figure}[h]
\begin{center}
\includegraphics[width=0.26\textwidth]{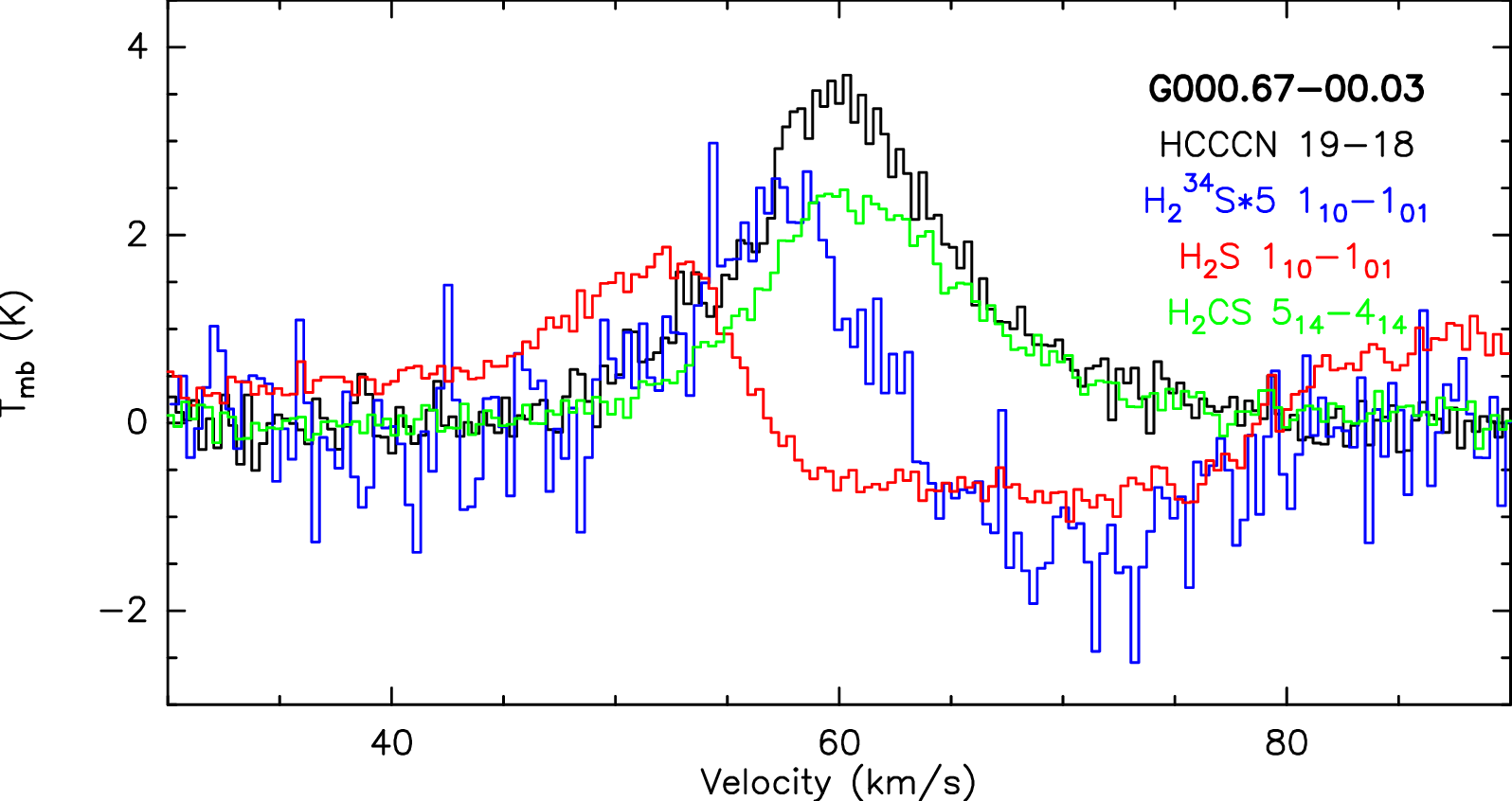}
\includegraphics[width=0.26\textwidth]{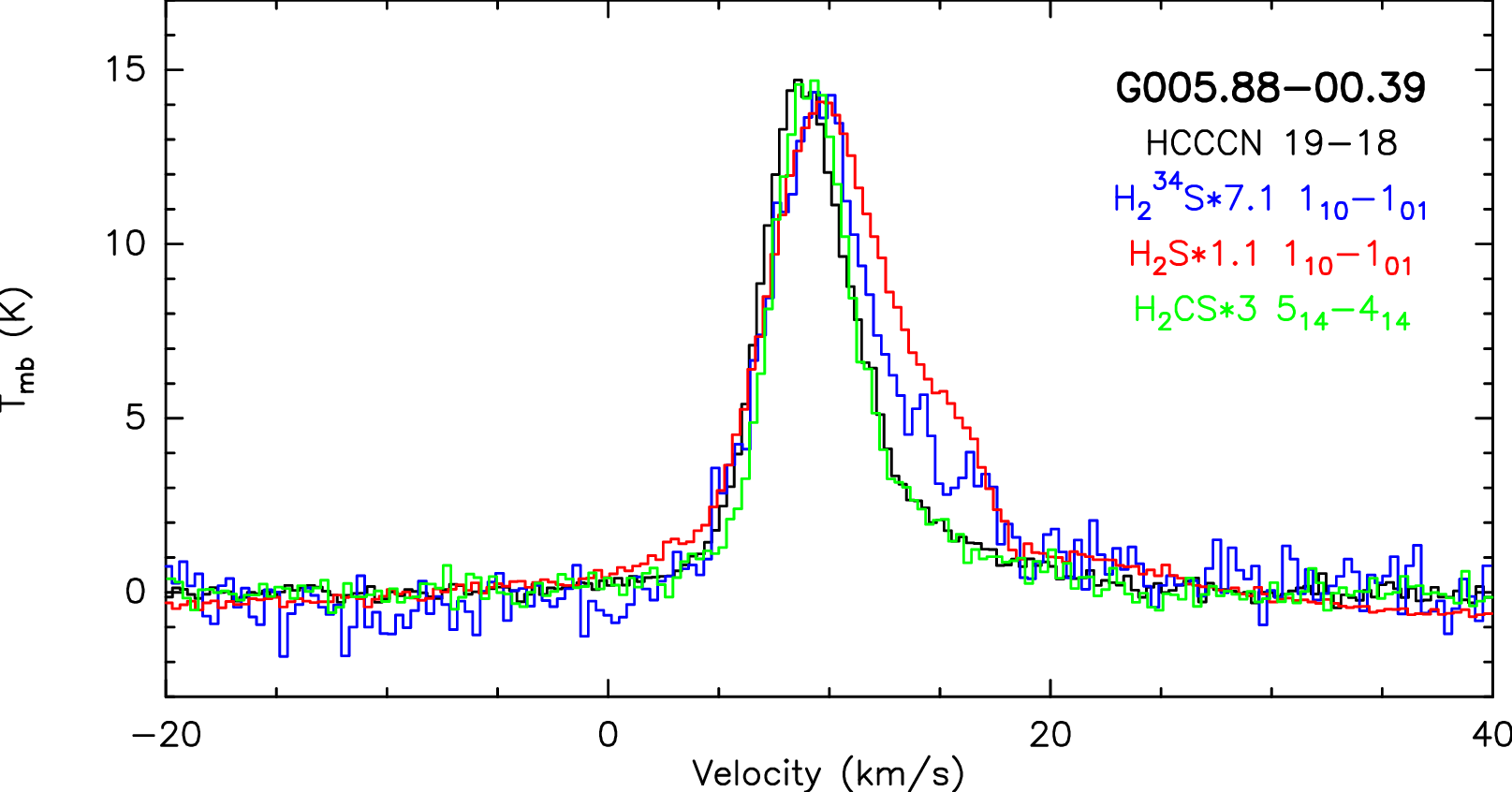}
\includegraphics[width=0.26\textwidth]{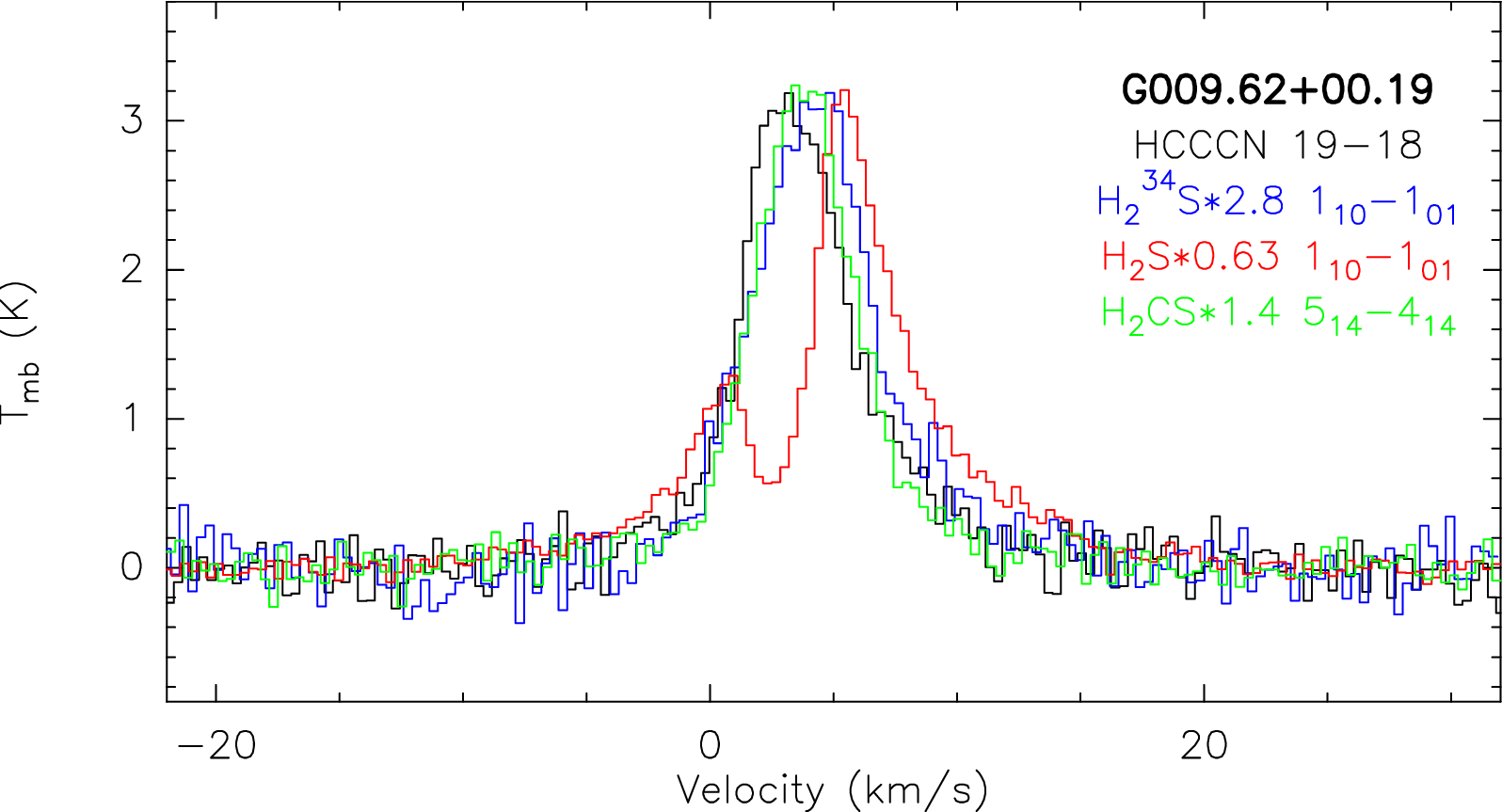}
\includegraphics[width=0.26\textwidth]{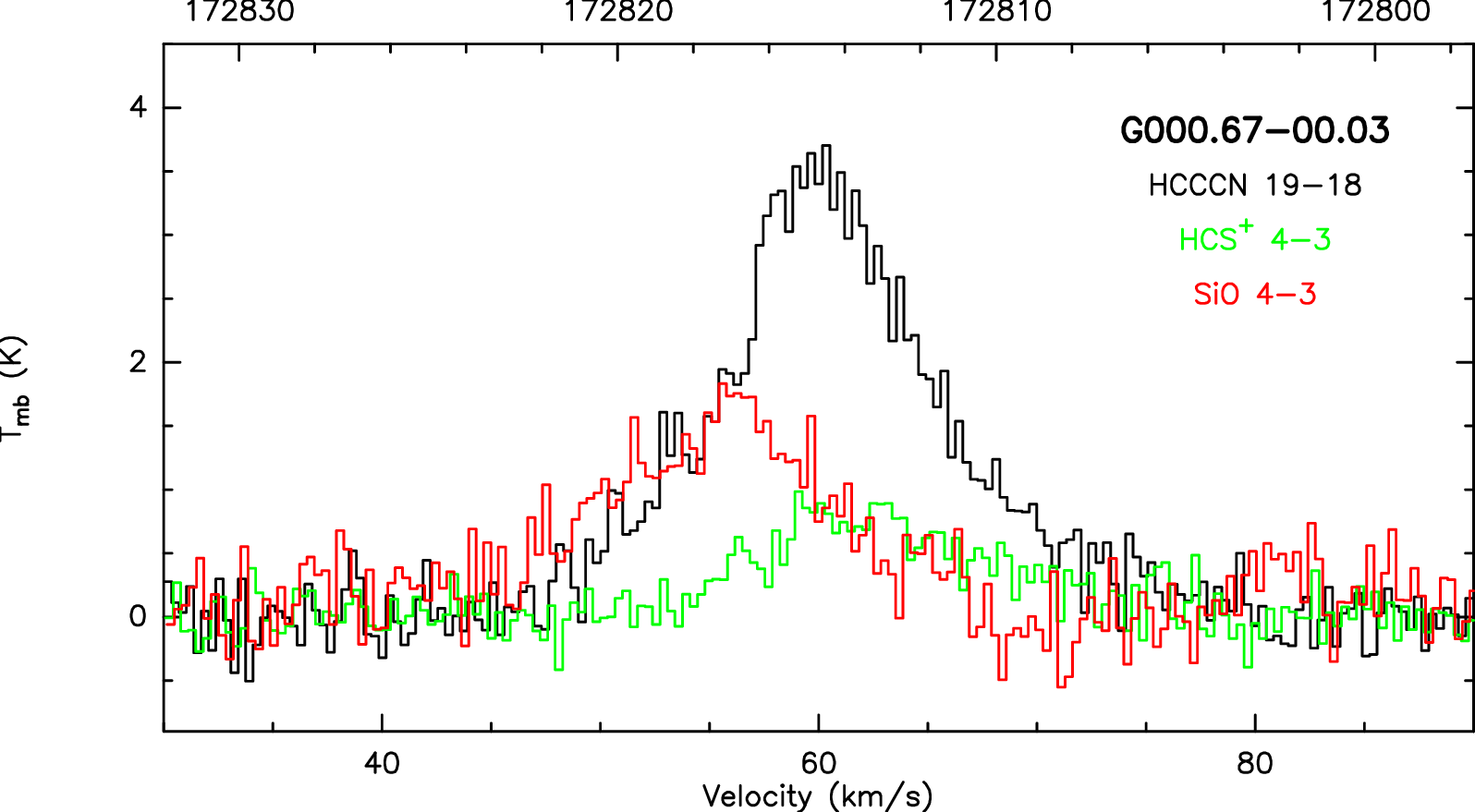}
\includegraphics[width=0.26\textwidth]{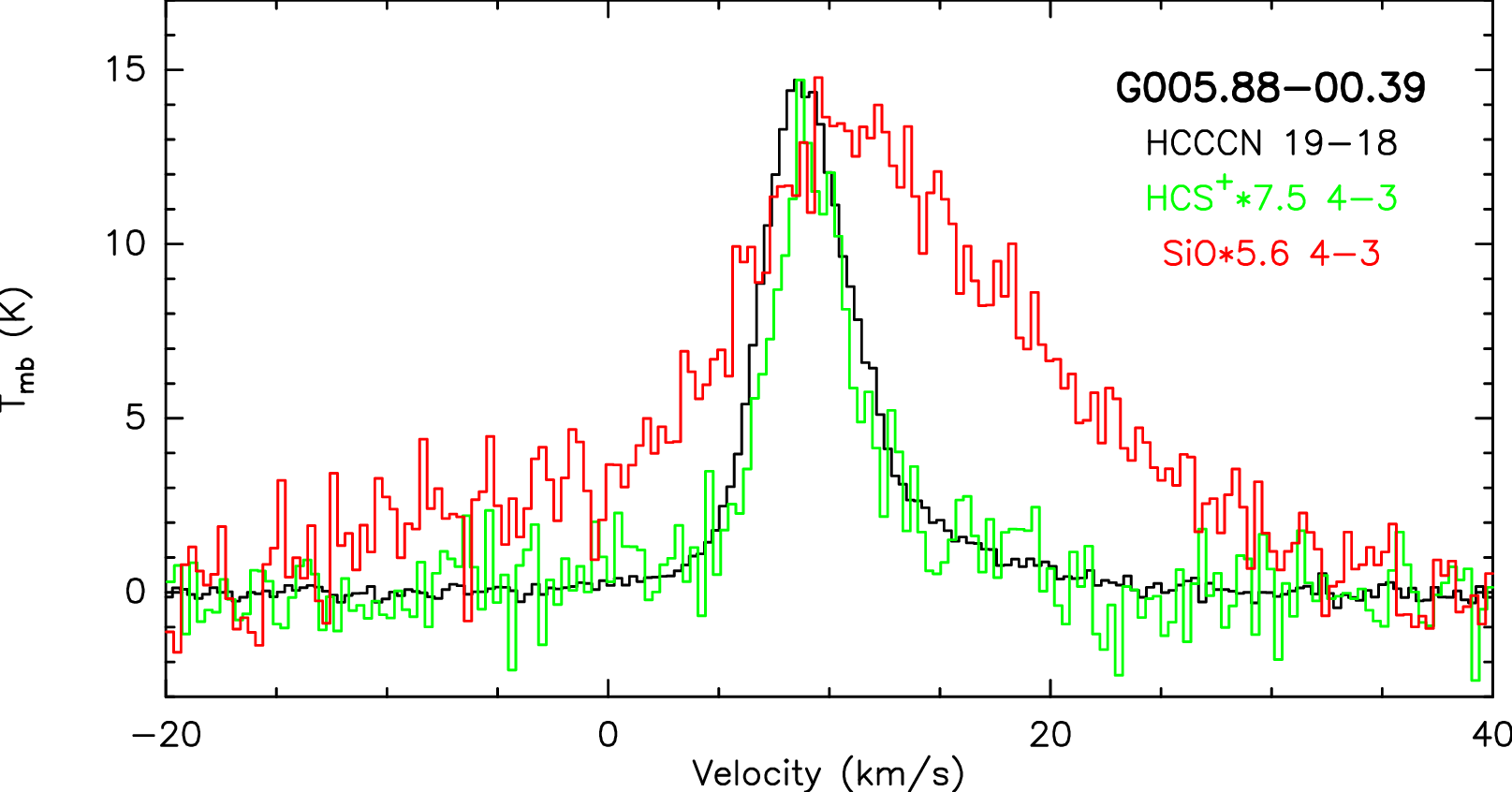}
\includegraphics[width=0.26\textwidth]{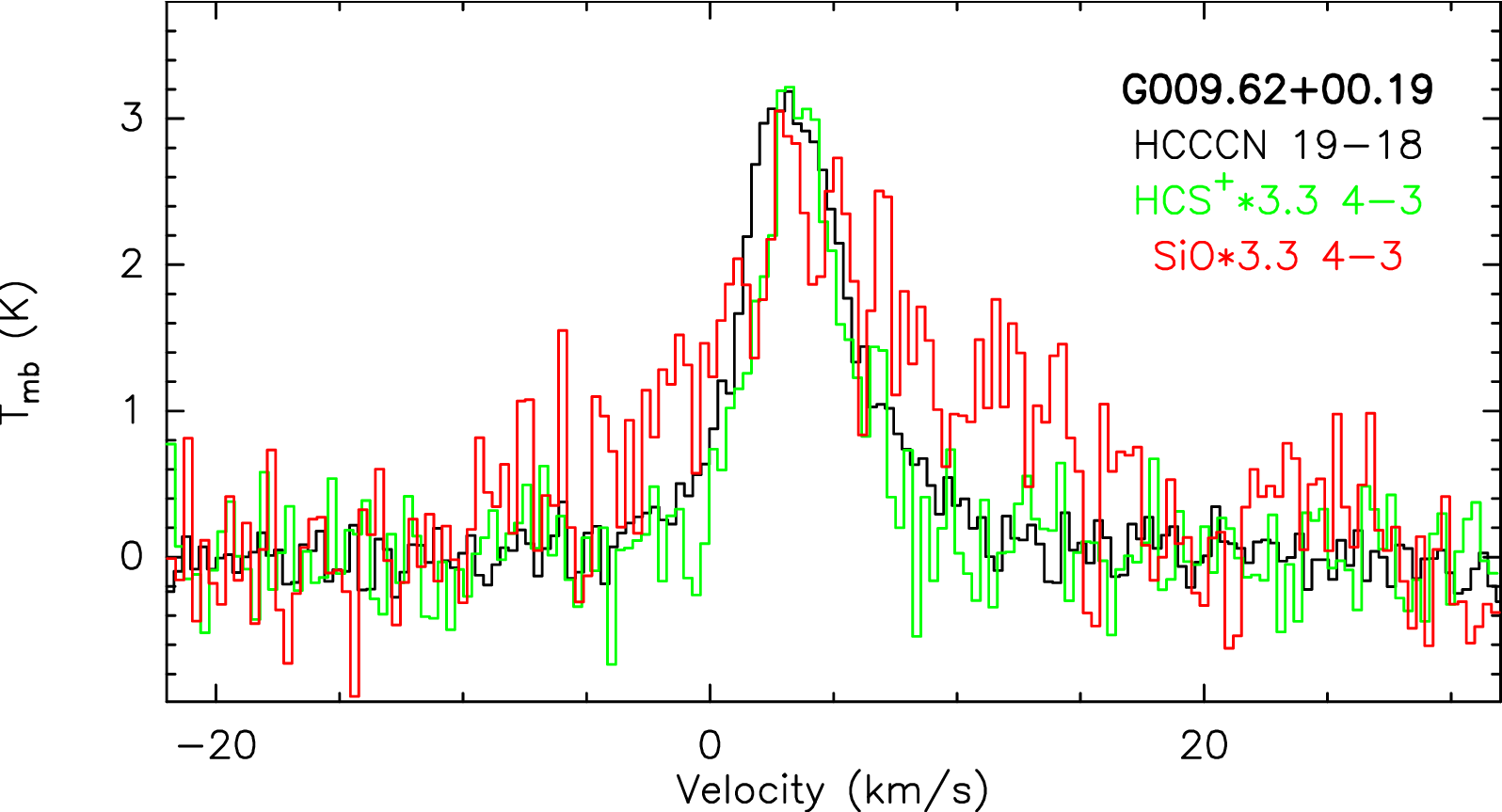}
\caption{The same as Figure 1 for more sources. More figures are presented in Zenodo. \url{https://doi.org/10.5281/zenodo.13937627}}\label{figA1}
\end{center}
\end{figure}

\clearpage

\section{Beam averaged column densities.} 
\begin{table}[h] 
\centering
\begin{threeparttable}
\setlength{\tabcolsep}{0.06in}
\caption{Beam averaged column densities.} \label{table:density}

\begin{tabular}{cccccccccc}
\hline
\hline

Source       & $^{b}$N(H$_{2}$S)     & $^{c}$N(H$_{2}$S)     & N(H$_{2}$$^{34}$S)    & N(H$_{2}$CS)        & N(HCS$^{+}$)           &N(HC$_{3}$N)         & N(SiO)              & N(C$^{18}$O)  & N(H$_{2}$)           \\
             & 10$^{14}$ cm$^{-2}$ &10$^{14}$ cm$^{-2}$ & 10$^{13}$ cm$^{-2}$ & 10$^{13}$ cm$^{-2}$ & 10$^{12}$ cm$^{-2}$  & 10$^{14}$ cm$^{-2}$ & 10$^{12}$ cm$^{-2}$ & 10$^{16}$ cm$^{-2}$  &  10$^{22}$ cm$^{-2}$  \\
             
\hline

G000.67-00.03	&	$^{a}$39.7$\pm$0.9		&	7.31$\pm$0.65	&	4.39$\pm$0.39	&	55.3$\pm$0.20	&	34.8$\pm$1.9	&	10.7 	$\pm$	0.17 	&	62.8$\pm$1.1	&	11.0$\pm$0.04	&	48.0$\pm$0.18	\\
G005.88-00.39	&		29.1$\pm$0.2		&	22.9$\pm$0.41	&	11.2$\pm$0.20	&	39.5$\pm$0.61	&	29.2$\pm$1.1	&	20.4 	$\pm$	0.26 	&	92.5$\pm$1.4	&	3.76$\pm$0.02	&	16.4$\pm$0.09	\\
G009.62+00.19	&	$^{a}$17.7$\pm$0.1		&	12.3$\pm$0.31	&	6.49$\pm$0.16	&	20.8$\pm$0.27	&	14.6$\pm$0.6	&	4.47 	$\pm$	0.09 	&	26.0$\pm$1.0	&	4.77$\pm$0.01	&	20.8$\pm$0.05	\\
G010.47+00.02	&	$^{a}$45.4$\pm$0.3		&	26.5$\pm$0.42	&	15.0$\pm$0.24	&	65.3$\pm$0.28	&	60.4$\pm$1.8	&	11.8 	$\pm$	0.07 	&	31.0$\pm$0.74	&	3.91$\pm$0.02	&	17.1$\pm$0.08	\\
G010.62-00.38	&	$^{a}$49.6$\pm$0.2		&	35.0$\pm$0.31	&	18.3$\pm$0.16	&	62.1$\pm$0.14	&	49.9$\pm$0.7	&	10.0 	$\pm$	0.09 	&	34.9$\pm$0.52	&	9.37$\pm$0.01	&	40.9$\pm$0.06	\\
G011.49-01.48	&		1.11$\pm$0.02		&	0.36$\pm$0.16	&	0.17$\pm$0.07	&	2.43$\pm$0.23	&	2.55$\pm$0.62	&	0.87 	$\pm$	0.07 	&	$\leq$1.07	&	0.62$\pm$0.01	&	2.72$\pm$0.05	\\
G011.91-00.61	&	$^{a}$4.72$\pm$0.20		&	3.54$\pm$0.14	&	1.75$\pm$0.07	&	5.88$\pm$0.14	&	3.72$\pm$0.24	&	1.40 	$\pm$	0.03 	&	8.19$\pm$0.24	&	1.27$\pm$0.01	&	5.55$\pm$0.04	\\
G012.80-00.20	&	$^{a}$8.44$\pm$0.03		&	6.50$\pm$0.12	&	3.17$\pm$0.06	&	21.4$\pm$0.13	&	20.6$\pm$0.2	&	4.77 	$\pm$	0.02 	&	7.50$\pm$0.15	&	7.30$\pm$0.01	&	31.9$\pm$0.03	\\
G012.88+00.48	&	$^{a}$10.1$\pm$0.08		&	7.05$\pm$0.10	&	3.39$\pm$0.05	&	14.5$\pm$0.11	&	10.7$\pm$0.2	&	1.85 	$\pm$	0.08 	&	8.39$\pm$0.18	&	3.54$\pm$0.01	&	15.4$\pm$0.03	\\
G012.90-00.26	&	$^{a}$8.34$\pm$0.07		&	6.38$\pm$0.15	&	3.07$\pm$0.07	&	10.8$\pm$0.15	&	6.31$\pm$0.32	&	2.40 	$\pm$	0.04 	&	9.61$\pm$0.34	&	3.86$\pm$0.01	&	16.9$\pm$0.05	\\
G014.33-00.64	&	$^{a}$8.29$\pm$0.13		&	6.80$\pm$0.33	&	3.12$\pm$0.15	&	21.6$\pm$0.37	&	14.4$\pm$0.7	&	3.00 	$\pm$	0.10 	&	29.4$\pm$1.4	&	1.32$\pm$0.01	&	5.75$\pm$0.04	\\
G015.03-00.67	&	$^{a}$2.54$\pm$0.02		&	2.15$\pm$0.20	&	1.01$\pm$0.09	&	2.72$\pm$0.09	&	2.42$\pm$0.26	&	2.05 	$\pm$	0.03 	&	$\leq$0.48	&	0.86$\pm$0.01	&	3.77$\pm$0.03	\\
G016.58-00.05	&	$^{a}$5.58$\pm$0.05		&	4.15$\pm$0.13	&	2.06$\pm$0.06	&	8.96$\pm$0.11	&	5.84$\pm$0.23	&	1.25 	$\pm$	0.03 	&	8.27$\pm$0.30	&	2.67$\pm$0.01	&	11.6$\pm$0.03	\\
G023.00-00.41	&	$^{a}$4.94$\pm$0.06		&	3.61$\pm$0.13	&	1.83$\pm$0.06	&	4.82$\pm$0.11	&	3.08$\pm$0.27	&	1.32 	$\pm$	0.03 	&	5.92$\pm$0.23	&	3.39$\pm$0.01	&	14.8$\pm$0.04	\\
G023.44-00.18	&		2.98$\pm$0.05		&	2.15$\pm$0.08	&	1.12$\pm$0.04	&	6.39$\pm$0.12	&	4.59$\pm$0.22	&	0.74 	$\pm$	0.03 	&	9.51$\pm$0.23	&	4.31$\pm$0.01	&	18.8$\pm$0.06	\\
G027.36-00.16	&		8.09$\pm$0.06		&	5.86$\pm$0.11	&	3.03$\pm$0.05	&	16.6$\pm$0.15	&	7.94$\pm$0.25	&	2.35 	$\pm$	0.04 	&	18.7$\pm$0.33	&	2.32$\pm$0.01	&	10.2$\pm$0.03	\\
G028.86+00.06	&		6.44$\pm$0.05		&	4.60$\pm$0.09	&	2.37$\pm$0.05	&	7.34$\pm$0.09	&	6.05$\pm$0.16	&	1.08 	$\pm$	0.02 	&	3.05$\pm$0.16	&	2.73$\pm$0.01	&	11.9$\pm$0.03	\\
G029.95-00.01	&		21.8$\pm$0.1		&	16.0$\pm$0.25	&	8.07$\pm$0.12	&	19.4$\pm$0.24	&	13.6$\pm$0.6	&	3.85 	$\pm$	0.10 	&	15.1$\pm$0.67	&	3.51$\pm$0.01	&	15.3$\pm$0.04	\\
G031.28+00.06	&	$^{a}$5.67$\pm$0.12		&	4.04$\pm$0.19	&	1.00$\pm$0.09	&	13.6$\pm$0.17	&	9.98$\pm$0.44	&	1.40 	$\pm$	0.05 	&	6.76$\pm$0.45	&	2.65$\pm$0.01	&	11.6$\pm$0.04	\\
G031.58+00.07	&	$^{a}$4.72$\pm$0.05		&	3.50$\pm$0.16	&	1.74$\pm$0.08	&	9.12$\pm$0.15	&	5.19$\pm$0.24	&	0.97 	$\pm$	0.03 	&	6.76$\pm$0.22	&	2.37$\pm$0.01	&	10.3$\pm$0.04	\\
G032.04+00.05	&	$^{a}$7.16$\pm$0.05		&	5.33$\pm$0.12	&	2.66$\pm$0.06	&	10.9$\pm$0.11	&	8.35$\pm$0.29	&	1.67 	$\pm$	0.03 	&	16.1$\pm$0.23	&	2.21$\pm$0.01	&	9.65$\pm$0.03	\\
G034.39+00.22	&		0.98$\pm$0.01		&	0.83$\pm$0.06	&	0.38$\pm$0.03	&	2.23$\pm$0.06	&	1.30$\pm$0.10	&	0.27 	$\pm$	0.01 	&	3.78$\pm$0.12	&	1.62$\pm$0.01	&	7.08$\pm$0.02	\\
G035.02+00.34	&	$^{a}$6.27$\pm$0.03		&	5.01$\pm$0.10	&	2.36$\pm$0.05	&	8.33$\pm$0.08	&	7.16$\pm$0.17	&	1.76 	$\pm$	0.02 	&	5.47$\pm$0.19	&	1.59$\pm$0.01	&	6.96$\pm$0.03	\\
G035.19-00.74	&	$^{a}$6.17$\pm$0.04		&	5.06$\pm$0.22	&	2.37$\pm$0.10	&	13.2$\pm$0.13	&	8.57$\pm$0.32	&	2.73 	$\pm$	0.03 	&	5.78$\pm$0.20	&	1.97$\pm$0.01	&	8.59$\pm$0.05	\\
G035.20-01.73	&		4.24$\pm$0.02		&	3.37$\pm$0.14	&	1.62$\pm$0.07	&	4.22$\pm$0.09	&	3.65$\pm$0.17	&	0.35 	$\pm$	0.03 	&	2.69$\pm$0.23	&	1.05$\pm$0.01	&	4.58$\pm$0.02	\\
G037.43+01.51	&		1.89$\pm$0.01		&	1.62$\pm$0.07	&	0.75$\pm$0.03	&	4.27$\pm$0.09	&	3.19$\pm$0.17	&	0.62 	$\pm$	0.02 	&	3.54$\pm$0.15	&	1.44$\pm$0.01	&	6.28$\pm$0.02	\\
G043.16+00.01	&	$^{a}$18.3$\pm$0.09		&	15.5$\pm$0.22	&	7.04$\pm$0.10	&	25.3$\pm$0.17	&	23.9$\pm$0.5	&	7.10 	$\pm$	0.06 	&	37.6$\pm$0.36	&	5.67$\pm$0.01	&	24.8$\pm$0.04	\\
G043.79-00.12	&	$^{a}$5.42$\pm$0.03		&	4.28$\pm$0.15	&	2.07$\pm$0.07	&	10.0$\pm$0.11	&	8.99$\pm$0.25	&	1.14 	$\pm$	0.03 	&	8.57$\pm$0.20	&	2.23$\pm$0.01	&	9.74$\pm$0.04	\\
G049.48-00.36	&		16.3$\pm$0.09		&	13.3$\pm$0.19	&	6.28$\pm$0.09	&	26.9$\pm$0.06	&	11.4$\pm$0.5	&	8.16 	$\pm$	0.04 	&	27.1$\pm$0.52	&	3.64$\pm$0.02	&	15.9$\pm$0.08	\\
G049.48-00.38	&	$^{a}$43.2$\pm$0.1		&	33.2$\pm$0.25	&	15.8$\pm$0.12	&	125$\pm$0.16	&	58.2$\pm$1.0	&	15.3 	$\pm$	0.09 	&	64.3$\pm$1.5	&	6.89$\pm$0.03	&	30.1$\pm$0.11	\\
G059.78+00.06	&		3.09$\pm$0.02		&	2.60$\pm$0.07	&	1.18$\pm$0.03	&	6.00$\pm$0.07	&	5.65$\pm$0.17	&	0.63 	$\pm$	0.02 	&	0.93$\pm$0.14	&	0.74$\pm$0.01	&	3.25$\pm$0.02	\\
G069.54-00.97	&	$^{a}$2.79$\pm$0.02		&	2.40$\pm$0.12	&	1.08$\pm$0.05	&	5.45$\pm$0.08	&	4.04$\pm$0.15	&	1.45 	$\pm$	0.02 	&	4.83$\pm$0.17	&	1.79$\pm$0.01	&	7.83$\pm$0.02	\\
G075.76+00.33	&	$^{a}$3.84$\pm$0.01		&	3.44$\pm$0.11	&	1.53$\pm$0.05	&	6.91$\pm$0.05	&	5.34$\pm$0.14	&	1.04 	$\pm$	0.01 	&	2.01$\pm$0.10	&	0.91$\pm$0.01	&	3.97$\pm$0.03	\\
G078.12+03.63	&		1.67$\pm$0.01		&	1.51$\pm$0.10	&	0.67$\pm$0.05	&	3.92$\pm$0.12	&	2.89$\pm$0.23	&	1.66 	$\pm$	0.03 	&	9.88$\pm$0.44	&	0.63$\pm$0.01	&	2.74$\pm$0.02	\\
G081.75+00.59	&		1.85$\pm$0.03		&	1.63$\pm$0.06	&	0.72$\pm$0.03	&	3.70$\pm$0.05	&	3.24$\pm$0.13	&	0.78 	$\pm$	0.01 	&	2.71$\pm$0.17	&	1.71$\pm$0.01	&	7.48$\pm$0.02	\\
G081.87+00.78	&		7.59$\pm$0.04		&	6.72$\pm$0.12	&	2.99$\pm$0.05	&	17.5$\pm$0.12	&	10.3$\pm$0.3	&	4.19 	$\pm$	0.03 	&	23.7$\pm$0.32	&	2.61$\pm$0.01	&	11.4$\pm$0.04	\\
G092.67+03.07	&		2.93$\pm$0.02		&	2.60$\pm$0.07	&	1.15$\pm$0.03	&	10.3$\pm$0.07	&	7.22$\pm$0.13	&	1.43 	$\pm$	0.02 	&	15.5$\pm$0.21	&	0.66$\pm$0.01	&	2.89$\pm$0.02	\\
G109.87+02.11	&	$^{a}$3.39$\pm$0.01		&	3.09$\pm$0.13	&	1.36$\pm$0.06	&	3.03$\pm$0.06	&	2.38$\pm$0.18	&	2.34 	$\pm$	0.03 	&	8.56$\pm$0.28	&	2.36$\pm$0.01	&	10.3$\pm$0.02	\\
G111.54+00.77	&	$^{a}$4.42$\pm$0.03		&	4.13$\pm$0.06	&	1.76$\pm$0.02	&	7.13$\pm$0.05	&	5.47$\pm$0.10	&	0.81 	$\pm$	0.01 	&	4.96$\pm$0.10	&	1.48$\pm$0.01	&	6.45$\pm$0.02	\\
G121.29+00.65	&	$^{a}$1.54$\pm$0.01		&	1.41$\pm$0.06	&	0.62$\pm$0.03	&	3.85$\pm$0.03	&	3.28$\pm$0.07	&	0.50 	$\pm$	0.01 	&	2.34$\pm$0.08	&	0.87$\pm$0.01	&	3.80$\pm$0.01	\\
G123.06-06.30	&		2.13$\pm$0.01		&	2.01$\pm$0.04	&	0.84$\pm$0.02	&	7.98$\pm$0.04	&	5.93$\pm$0.09	&	0.73 	$\pm$	0.01 	&	10.4$\pm$0.11	&	0.72$\pm$0.01	&	3.16$\pm$0.02	\\
G133.94+01.06	&	$^{a}$9.99$\pm$0.04		&	8.85$\pm$0.15	&	3.74$\pm$0.06	&	22.5$\pm$0.12	&	17.8$\pm$0.2	&	2.66 	$\pm$	0.02 	&	15.6$\pm$0.16	&	1.86$\pm$0.01	&	8.12$\pm$0.03	\\
G168.06+00.82	&	$^{a}$0.64$\pm$0.03		&	0.72$\pm$0.25	&	0.26$\pm$0.08	&	0.71$\pm$0.09	&	0.67$\pm$0.22	&	0.17 	$\pm$	0.07 	&	$\leq$0.47	&	0.33$\pm$0.01	&	1.45$\pm$0.03	\\
G176.51+00.20	&	$^{a}$0.78$\pm$0.01		&	0.73$\pm$0.09	&	0.31$\pm$0.04	&	1.45$\pm$0.05	&	1.19$\pm$0.12	&	0.20 	$\pm$	0.01 	&	1.16$\pm$0.13	&	0.56$\pm$0.01	&	2.44$\pm$0.01	\\
G183.72-03.66	&		0.52$\pm$0.01		&	0.50$\pm$0.04	&	0.21$\pm$0.02	&	1.52$\pm$0.04	&	1.41$\pm$0.09	&	0.21 	$\pm$	0.01 	&	2.81$\pm$1.12	&	0.26$\pm$0.01	&	1.13$\pm$0.01	\\
G188.94+00.88	&		1.87$\pm$0.01		&	1.82$\pm$0.09	&	0.75$\pm$0.04	&	4.11$\pm$0.07	&	3.86$\pm$0.15	&	0.26 	$\pm$	0.01 	&	2.02$\pm$0.10	&	0.58$\pm$0.01	&	2.53$\pm$0.02	\\
G192.60-00.04	&		3.81$\pm$0.01		&	3.60$\pm$0.07	&	1.52$\pm$0.03	&	4.53$\pm$0.07	&	4.95$\pm$0.15	&	0.46 	$\pm$	0.02 	&	3.97$\pm$0.18	&	0.96$\pm$0.01	&	4.18$\pm$0.02	\\
G209.00-19.38	&		1.78$\pm$0.01		&	1.63$\pm$0.05	&	0.72$\pm$0.02	&	1.94$\pm$0.32	&	2.03$\pm$0.08	&	0.47 	$\pm$	0.01 	&	3.53$\pm$0.13	&	0.40$\pm$0.01	&	1.76$\pm$0.01	\\
G232.62+00.99	&		0.93$\pm$0.01		&	0.88$\pm$0.07	&	0.38$\pm$0.03	&	2.30$\pm$0.05	&	2.90$\pm$0.21	&	0.49 	$\pm$	0.02 	&	$\leq$0.33	&	0.63$\pm$0.01	&	2.75$\pm$0.02	\\
G211.59+01.05	&		0.82$\pm$0.01		&	0.82$\pm$0.03	&	0.33$\pm$0.01	&	1.34$\pm$0.03	&	1.35$\pm$0.07	&	0.15 	$\pm$	0.01 	&	2.06$\pm$0.07	&	0.47$\pm$0.01	&	2.06$\pm$0.01	\\
\hline

\end{tabular}
\begin{tablenotes}
\footnotesize
\item  Notes. $^{a}$ The flux of H$_{2}$S lines are obtained by "print area'".  $^{b}$  The beam averaged column densities of H$_{2}$S are calculated by corrected optical depths.  $^{c}$ The beam averaged column densities of H$_{2}$S are derived from those of H$_{2}$$^{34}$S with $^{32}$S/$^{34}$S ratios.
\end{tablenotes}
\end{threeparttable}
\end{table}

\newpage

\section{Line widths at half maximum.} 
\begin{table}[h]
\centering
\caption{ The line widths at half maximum in km s$^{-1}$  for the molecules.} \label{table:fwhm}

\begin{tabular}{cccccc}
\hline
\hline

Source       &            HC$_{3}$N   &H$_{2}$$^{34}$S     &  H$_{2}$CS        & HCS$^{+}$                    & SiO                     \\
             
\hline

G000.67-00.03 	&	12.0 $\pm$	0.24 	&	11.3 $\pm$	0.81 	&	12.28 $\pm$	0.25 	&	12.7 $\pm$	0.78 	&	13.0 	$\pm$	0.30 	\\
G005.88-00.39	&	5.23 	$\pm$	0.08 	&	6.38 	$\pm$	0.14 	&	4.62 	$\pm$	0.09 	&	4.86 	$\pm$	0.24 	&	27.5 	$\pm$	0.55 	\\
G009.62+00.19	&	5.34 	$\pm$	0.14 	&	5.98 	$\pm$	0.15 	&	4.94 	$\pm$	0.08 	&	4.34 	$\pm$	0.22 	&	26.5 	$\pm$	1.47 	\\
G010.47+00.02	&	9.64 	$\pm$	0.06 	&	6.46 	$\pm$	0.43 	&	8.37 	$\pm$	0.06 	&	7.06 	$\pm$	0.64 	&	14.9 	$\pm$	0.56 	\\
G010.62-00.38	&	7.49 	$\pm$	0.08 	&	6.57 	$\pm$	0.07 	&	7.03 	$\pm$	0.04 	&	7.13 	$\pm$	0.12 	&	9.96 	$\pm$	0.19 	\\
G011.49-01.48	&	1.93 	$\pm$	0.15 	&	0.66 	$\pm$	0.39 	&	2.40 	$\pm$	0.25 	&	2.50 	$\pm$	0.51 	&	undetected			\\
G011.91-00.61	&	6.71 	$\pm$	0.15 	&	6.51 	$\pm$	0.29 	&	5.68 	$\pm$	0.17 	&	4.64 	$\pm$	0.40 	&	13.1 	$\pm$	0.52 	\\
G012.80-00.20	&	5.38 	$\pm$	0.03 	&	5.57 	$\pm$	0.12 	&	5.22 	$\pm$	0.02 	&	5.38 	$\pm$	0.07 	&	7.09 	$\pm$	0.18 	\\
G012.88+00.48	&	4.11 	$\pm$	0.19 	&	4.14 	$\pm$	0.06 	&	3.71 	$\pm$	0.03 	&	3.34 	$\pm$	0.07 	&	11.9 	$\pm$	0.34 	\\
G012.90-00.26	&	4.77 	$\pm$	0.10 	&	4.71 	$\pm$	0.12 	&	4.51 	$\pm$	0.08 	&	4.09 	$\pm$	0.25 	&	11.8 	$\pm$	0.55 	\\
G014.33-00.64	&	3.26 	$\pm$	0.14 	&	3.00 	$\pm$	0.12 	&	2.98 	$\pm$	0.09 	&	2.83 	$\pm$	0.17 	&	18.4 	$\pm$	1.03 	\\
G015.03-00.67	&	2.63 	$\pm$	0.04 	&	3.85 	$\pm$	0.32 	&	2.37 	$\pm$	0.12 	&	2.25 	$\pm$	0.29 	&	undetected		\\
G016.58-00.05	&	4.69 	$\pm$	0.15 	&	3.82 	$\pm$	0.11 	&	4.06 	$\pm$	0.05 	&	3.53 	$\pm$	0.16 	&	9.62 	$\pm$	0.49 	\\
G023.00-00.41	&	6.32 	$\pm$	0.22 	&	8.14 	$\pm$	0.34 	&	4.99 	$\pm$	0.25 	&	5.71 	$\pm$	0.62 	&	17.2 	$\pm$	0.87 	\\
G023.44-00.18	&	4.16 	$\pm$	0.20 	&	3.76 	$\pm$	0.19 	&	4.24 	$\pm$	0.10 	&	4.27 	$\pm$	0.26 	&	14.7 	$\pm$	0.45 	\\
G027.36-00.16	&	7.46 	$\pm$	0.19 	&	6.28 	$\pm$	0.14 	&	5.43 	$\pm$	0.06 	&	4.78 	$\pm$	0.18 	&	13.0 	$\pm$	0.33 	\\
G028.86+00.06	&	4.28 	$\pm$	0.11 	&	4.21 	$\pm$	0.11 	&	3.39 	$\pm$	0.05 	&	2.98 	$\pm$	0.10 	&	7.11 	$\pm$	0.54 	\\
G029.95-00.01	&	6.25 	$\pm$	0.21 	&	4.67 	$\pm$	0.09 	&	3.95 	$\pm$	0.06 	&	3.69 	$\pm$	0.20 	&	10.8 	$\pm$	0.67 	\\
G031.28+00.06	&	3.75 	$\pm$	0.16 	&	3.93 	$\pm$	0.17 	&	3.98 	$\pm$	0.06 	&	3.25 	$\pm$	0.17 	&	7.36 	$\pm$	0.64 	\\
G031.58+00.07	&	3.36 	$\pm$	0.11 	&	3.73 	$\pm$	0.16 	&	3.39 	$\pm$	0.06 	&	3.08 	$\pm$	0.17 	&	7.09 	$\pm$	0.28 	\\
G032.04+00.05	&	5.79 	$\pm$	0.13 	&	6.24 	$\pm$	0.16 	&	5.32 	$\pm$	0.07 	&	5.23 	$\pm$	0.22 	&	12.2 	$\pm$	0.22 	\\
G034.39+00.22	&	2.95 	$\pm$	0.21 	&	3.29 	$\pm$	0.32 	&	3.00 	$\pm$	0.11 	&	1.86 	$\pm$	0.14 	&	4.37 	$\pm$	0.19 	\\
G035.02+00.34	&	5.87 	$\pm$	0.06 	&	4.58 	$\pm$	0.08 	&	4.47 	$\pm$	0.05 	&	4.14 	$\pm$	0.12 	&	9.56 	$\pm$	0.48 	\\
G035.19-00.74	&	4.73 	$\pm$	0.07 	&	5.08 	$\pm$	0.16 	&	4.58 	$\pm$	0.06 	&	3.86 	$\pm$	0.17 	&	8.63 	$\pm$	0.44 	\\
G035.20-01.73	&	3.48 	$\pm$	0.62 	&	3.73 	$\pm$	0.11 	&	2.96 	$\pm$	0.08 	&	2.93 	$\pm$	0.25 	&	11.1 	$\pm$	1.50 	\\
G037.43+01.51	&	2.86 	$\pm$	0.12 	&	2.63 	$\pm$	0.14 	&	2.89 	$\pm$	0.07 	&	2.65 	$\pm$	0.19 	&	4.31 	$\pm$	0.25 	\\
G043.16+00.01	&	14.4 	$\pm$	0.15 	&	16.7 	$\pm$	0.24 	&	14.5 	$\pm$	0.11 	&	15.8 	$\pm$	0.29 	&	14.4 	$\pm$	0.15 	\\
G043.79-00.12	&	5.94 	$\pm$	0.17 	&	5.95 	$\pm$	0.17 	&	5.48 	$\pm$	0.06 	&	5.50 	$\pm$	0.18 	&	8.78 	$\pm$	0.25 	\\
G049.48-00.36	&	6.73 	$\pm$	0.04 	&	7.21 	$\pm$	0.12 	&	7.89 	$\pm$	0.07 	&	6.58 	$\pm$	0.32 	&	11.4 	$\pm$	0.28 	\\
G049.48-00.38	&	9.05 	$\pm$	0.07 	&	8.46 	$\pm$	0.09 	&	7.30 	$\pm$	0.03 	&	5.89 	$\pm$	0.12 	&	10.7 	$\pm$	0.24 	\\
G059.78+00.06	&	2.14 	$\pm$	0.07 	&	1.92 	$\pm$	0.07 	&	2.03 	$\pm$	0.03 	&	2.16 	$\pm$	0.08 	&	4.64 	$\pm$	0.92 	\\
G069.54-00.97	&	4.20 	$\pm$	0.06 	&	3.57 	$\pm$	0.13 	&	3.39 	$\pm$	0.06 	&	3.09 	$\pm$	0.13 	&	5.56 	$\pm$	0.29 	\\
G075.76+00.33	&	3.93 	$\pm$	0.05 	&	3.32 	$\pm$	0.07 	&	3.53 	$\pm$	0.03 	&	3.35 	$\pm$	0.10 	&	5.48 	$\pm$	0.34 	\\
G078.12+03.63	&	4.09 	$\pm$	0.08 	&	3.96 	$\pm$	0.32 	&	4.93 	$\pm$	0.19 	&	3.64 	$\pm$	0.37 	&	29.0 	$\pm$	1.81 	\\
G081.75+00.59	&	2.16 	$\pm$	0.04 	&	2.16 	$\pm$	0.10 	&	2.07 	$\pm$	0.03 	&	1.95 	$\pm$	0.10 	&	10.7 	$\pm$	1.02 	\\
G081.87+00.78	&	5.49 	$\pm$	0.05 	&	4.38 	$\pm$	0.09 	&	4.43 	$\pm$	0.04 	&	4.03 	$\pm$	0.14 	&	7.73 	$\pm$	0.14 	\\
G092.67+03.07	&	3.15 	$\pm$	0.04 	&	2.99 	$\pm$	0.11 	&	3.46 	$\pm$	0.03 	&	2.91 	$\pm$	0.07 	&	19.2 	$\pm$	0.34 	\\
G109.87+02.11	&	4.00 	$\pm$	0.06 	&	4.60 	$\pm$	0.17 	&	3.38 	$\pm$	0.15 	&	3.14 	$\pm$	0.32 	&	14.2 	$\pm$	0.60 	\\
G111.54+00.77	&	4.24 	$\pm$	0.08 	&	4.30 	$\pm$	0.08 	&	3.89 	$\pm$	0.03 	&	3.79 	$\pm$	0.08 	&	6.86 	$\pm$	0.20 	\\
G121.29+00.65	&	2.60 	$\pm$	0.04 	&	2.75 	$\pm$	0.07 	&	2.59 	$\pm$	0.03 	&	2.36 	$\pm$	0.06 	&	3.79 	$\pm$	0.20 	\\
G123.06-06.30	&	3.20 	$\pm$	0.04 	&	3.57 	$\pm$	0.09 	&	3.64 	$\pm$	0.02 	&	3.42 	$\pm$	0.07 	&	7.79 	$\pm$	0.11 	\\
G133.94+01.06	&	4.49 	$\pm$	0.05 	&	5.28 	$\pm$	0.09 	&	3.93 	$\pm$	0.02 	&	3.81 	$\pm$	0.05 	&	6.52 	$\pm$	0.08 	\\
G168.06+00.82	&	4.66 	$\pm$	1.98 	&	1.23 	$\pm$	0.21 	&	1.34 	$\pm$	0.30 	&	0.83 	$\pm$	0.28 	&	undetected			\\
G176.51+00.20	&	1.89 	$\pm$	0.12 	&	2.98 	$\pm$	0.33 	&	2.07 	$\pm$	0.10 	&	1.60 	$\pm$	0.23 	&	14.8 	$\pm$	1.35 	\\
G183.72-03.66	&	1.95 	$\pm$	0.08 	&	2.47 	$\pm$	0.28 	&	2.18 	$\pm$	0.07 	&	2.38 	$\pm$	0.18 	&	8.55 	$\pm$	0.49 	\\
G188.94+00.88	&	2.49 	$\pm$	0.17 	&	3.02 	$\pm$	0.18 	&	2.63 	$\pm$	0.05 	&	2.95 	$\pm$	0.13 	&	4.00 	$\pm$	0.24 	\\
G192.60-00.04	&	3.72 	$\pm$	0.18 	&	3.16 	$\pm$	0.07 	&	2.56 	$\pm$	0.05 	&	2.03 	$\pm$	0.08 	&	9.18 	$\pm$	0.62 	\\
G209.00-19.38	&	2.21 	$\pm$	0.05 	&	3.13 	$\pm$	0.11 	&	2.25 	$\pm$	0.04 	&	2.35 	$\pm$	0.11 	&	18.8 	$\pm$	0.97 	\\
G232.62+00.99	&	2.12 	$\pm$	0.09 	&	2.40 	$\pm$	0.26 	&	1.96 	$\pm$	0.05 	&	2.91 	$\pm$	0.20 	&	undetected		\\
G211.59+01.05	&	3.06 	$\pm$	0.16 	&	3.27 	$\pm$	0.12 	&	3.40 	$\pm$	0.08 	&	2.58 	$\pm$	0.28 	&	6.87 	$\pm$	0.29 	\\

\hline

\end{tabular}
\end{table}

\clearpage

\section{$^{32}$S/$^{34}$S ratios, $\tau$$_{H_{2}S}$ and abundances.}
\begin{table}[h]
\centering
\caption{The information of $^{32}$S/$^{34}$S ratios, $\tau$$_{H_{2}S}$ and abundances.} \label{tableA1}

\begin{tabular}{ccccccc}
\hline
\multirow{2}{*}{Source} & \multirow{2}{*}{$^{32}$S/$^{34}$S} &    \multirow{2}{*}{$\tau_{H_{2}S}$}    & \multicolumn{4}{c}{Abundance}                                              \\ \cmidrule(r){4-7}
                        &              &       &          H$_{2}$S (10$^{-9}$) & H$_{2}$CS (10$^{-10}$) & HCS$^{+}$ (10$^{-11}$) & SiO (10$^{-11}$) \\
\hline
G000.67-00.03            & 16.65          &        6.60            & 1.52        $\pm$ 0.13       & 11.5          $\pm$         0.06        & 7.25           $\pm$         0.40        & 13.1   $\pm$   0.24  \\
G005.88-00.39            & 20.37           &        3.13           & 13.9        $\pm$  0.26       & 24.0          $\pm$         0.39        & 17.8          $\pm$         0.68        & 56.3    $\pm$   0.88  \\
G009.62+00.19            & 18.91          &        5.77            & 5.89         $\pm$  0.15       & 9.98           $\pm$         0.13        & 7.02           $\pm$         0.29        & 12.5   $\pm$   0.51  \\
G010.47+00.02            & 17.67          &         11.2           & 15.5          $\pm$ 0.25       & 38.3          $\pm$         0.23        & 35.4          $\pm$         1.07        & 18.2    $\pm$   0.44  \\
G010.62-00.38            & 19.13           &       5.46            & 8.55        $\pm$  0.08       & 15.2          $\pm$         0.04        & 12.2          $\pm$         0.18        & 8.54    $\pm$   0.13  \\
G011.49-01.48            & 21.68            &       0.72           & 6.37      $\pm$  0.25       & 10.6          $\pm$         0.27        & 6.70           $\pm$         0.44        & 14.8   $\pm$   0.45  \\
G011.91-00.61            & 20.22            &        5.21          & 2.04         $\pm$  0.04       & 6.72           $\pm$         0.04        & 6.46           $\pm$         0.07        & 2.4    $\pm$   0.05  \\
G012.80-00.20            & 20.52           &        4.61           & 4.56        $\pm$  0.07       & 9.39           $\pm$         0.07        & 6.91           $\pm$         0.12        & 5.43    $\pm$   0.12  \\
G012.88+00.48            & 20.81          &        10.4            & 3.79         $\pm$  0.09       & 6.41           $\pm$         0.09        & 3.75           $\pm$         0.19        & 5.70    $\pm$   0.20  \\
G012.90-00.26            & 20.81           &         5.64          & 11.8        $\pm$  0.59       & 37.5          $\pm$         0.71        & 25.0          $\pm$         1.24        & 51.2   $\pm$   2.37  \\
G014.33-00.64            & 21.76           &         4.39         & 3.56         $\pm$  0.11       & 7.69           $\pm$         0.10        & 5.01           $\pm$         0.20        & 7.10    $\pm$   0.26  \\
G015.03-00.67            & 21.17          &          1.47          & 2.44          $\pm$  0.09       & 3.25           $\pm$         0.08        & 2.07           $\pm$         0.18        & 3.99    $\pm$   0.16  \\
G016.58-00.05            & 20.15            &         5.44         & 1.14         $\pm$  0.04       & 3.39           $\pm$         0.06        & 2.44           $\pm$         0.12        & 5.05    $\pm$   0.12  \\
G023.00-00.41            & 19.79           &         5.38          & 5.78        $\pm$  0.10       & 16.4          $\pm$         0.15        & 7.82           $\pm$         0.24        & 18.4    $\pm$   0.33  \\
G023.44-00.18            & 19.20           &         4.49          & 3.85          $\pm$  0.08       & 6.15           $\pm$         0.07        & 5.06           $\pm$         0.13        & 2.56    $\pm$   0.13  \\
G027.36-00.16            & 19.35           &         4.69          & 10.5         $\pm$ 0.16       & 12.6          $\pm$         0.16        & 8.87           $\pm$         0.39        & 9.86    $\pm$   0.44  \\
G028.86+00.06            & 19.42          &          5.68          & 3.49         $\pm$  0.16       & 11.8          $\pm$         0.15        & 8.63           $\pm$         0.39        & 5.85    $\pm$   0.39  \\
G029.95-00.01            & 19.86            &        5.36          & 3.39          $\pm$  0.16       & 8.83           $\pm$         0.15        & 5.03           $\pm$         0.24        & 6.55    $\pm$   0.21  \\
G031.28+00.06            & 20.30          &         7.96           & 5.52          $\pm$  0.13       & 11.3          $\pm$         0.12        & 8.65           $\pm$         0.30        & 16.7   $\pm$   0.24  \\
G031.58+00.07            & 20.08         &           5.35          & 1.17         $\pm$  0.09       & 3.15           $\pm$         0.09        & 1.84           $\pm$         0.14        & 5.34    $\pm$   0.17  \\
G032.04+00.05            & 20.00           &       5.05            & 7.19         $\pm$  0.15       & 12.0          $\pm$         0.12        & 10.3          $\pm$         0.25        & 7.86    $\pm$   0.27  \\
G034.39+00.22            & 21.68           &        2.45           & 5.89          $\pm$  0.26       & 15.3          $\pm$         0.17        & 9.98           $\pm$         0.37        & 6.73    $\pm$   0.24  \\
G035.02+00.34            & 21.25           &       4.56            & 7.36          $\pm$ 0.30       & 9.22           $\pm$         0.19        & 7.99           $\pm$        0.37        & 5.87    $\pm$   0.51  \\
G035.19-00.74            & 21.32            &       3.38           & 2.57          $\pm$ 0.11       & 6.80           $\pm$         0.14        & 5.09           $\pm$         0.28        & 5.64    $\pm$   0.25  \\
G035.20-01.73            & 20.81           &         3.72         & 6.26         $\pm$ 0.09       & 10.2          $\pm$        0.07        & 9.66           $\pm$         0.18        & 15.2    $\pm$   0.15  \\
G037.43+01.51            & 21.54           &        1.80           & 4.40         $\pm$ 0.16       & 10.3          $\pm$         0.12        & 9.22           $\pm$         0.26        & 8.80    $\pm$   0.21  \\
G043.16+00.01            & 22.05          &        3.31            & 8.35         $\pm$ 0.13       & 16.9          $\pm$        0.09        & 7.15           $\pm$        0.29        & 17.1   $\pm$   0.34  \\
G043.79-00.12            & 20.66            &       3.65           & 11.1        $\pm$ 0.09       & 41.5          $\pm$         0.17        & 19.3          $\pm$         0.34        & 21.4   $\pm$   0.52  \\
G049.48-00.36            & 21.10           &        3.12           & 8.00         $\pm$ 0.22       & 18.5          $\pm$         0.24        & 17.4          $\pm$         0.52        & 2.86    $\pm$   0.42  \\
G049.48-00.38            & 21.10            &       6.03           & 3.07          $\pm$ 0.15       & 6.96           $\pm$         0.10        & 5.16          $\pm$         0.20        & 6.16    $\pm$   0.22  \\
G059.78+00.06            & 21.98           &        3.66           & 8.67          $\pm$ 0.28       & 17.4          $\pm$         0.17        & 13.4           $\pm$         0.36        & 5.07    $\pm$   0.25  \\
G069.54-00.97            & 22.19            &       2.86           & 5.51        $\pm$ 0.37       & 14.3          $\pm$         0.44        & 10.52          $\pm$         0.85        & 36.0    $\pm$   1.62  \\
G075.76+00.33            & 22.49           &         1.64          & 2.18          $\pm$ 0.08       & 4.94           $\pm$         0.06        & 4.33           $\pm$         0.17        & 3.61   $\pm$  0.22  \\
G078.12+03.63            & 22.41          &       0.99             & 5.90          $\pm$ 0.11       & 15.3          $\pm$         0.11        & 9.01           $\pm$         0.26        & 20.8   $\pm$   0.29  \\
G081.75+00.59            & 22.49          &       2.46             & 8.99          $\pm$         0.26       & 35.7          $\pm$         0.32        & 25.0          $\pm$         0.48        & 53.5    $\pm$   0.79  \\
G081.87+00.78            & 22.49          &        2.20            & 3.00          $\pm$         0.13       & 2.94           $\pm$         0.06        & 2.32           $\pm$         0.17        & 8.32    $\pm$   0.27  \\
G092.67+03.07            & 22.71          &        2.57            & 6.41          $\pm$         0.09       & 11.1          $\pm$         0.08        & 8.48           $\pm$         0.16        & 7.69    $\pm$   0.16  \\
G109.87+02.11            & 22.78           &        1.39           & 3.71          $\pm$         0.16       & 10.1          $\pm$         0.09        & 8.64           $\pm$         0.18        & 6.17    $\pm$   0.22  \\
G111.54+00.77            & 23.51          &          1.72          & 6.37          $\pm$         0.13       & 25.3          $\pm$         0.18        & 18.8          $\pm$         0.31        & 33.0    $\pm$   0.38  \\
G121.29+00.65            & 22.92           &        1.49           & 10.9         $\pm$         0.19       & 27.7          $\pm$         0.17        & 21.9          $\pm$         0.25        & 19.2    $\pm$   0.21  \\
G123.06-06.30            & 23.87           &         2.02          & 2.99          $\pm$         0.38       & 5.94          $\pm$         0.20        & 4.88           $\pm$         0.51        & 4.76    $\pm$   0.52  \\
G133.94+01.06            & 23.65          &         4.72           & 4.37          $\pm$         0.36       & 13.4          $\pm$         0.35        & 12.4          $\pm$         0.78        & 24.8   $\pm$   1.01  \\
G168.06+00.82            & 28.11           &         1.06          & 7.17          $\pm$         0.35       & 16.2          $\pm$         0.29        & 15.2          $\pm$         0.59        & 7.99    $\pm$   0.41  \\
G176.51+00.20            & 23.29           &         1.07          & 8.62          $\pm$         0.16       & 10.9          $\pm$         0.18        & 11.9          $\pm$         0.37        & 9.50    $\pm$   0.44  \\
G183.72-03.66            & 23.80            &         1.40         & 9.29          $\pm$         0.27       & 11.0          $\pm$         1.80        & 11.6          $\pm$         0.46        & 20.1   $\pm$   0.76  \\
G188.94+00.88            & 24.09            &        1.06          & 4.00          $\pm$         0.14       & 6.52           $\pm$         0.14        & 6.6           $\pm$         0.35        & 9.99    $\pm$  0.33  \\
G192.60-00.04            & 23.73           &         1.62          & 1.32          $\pm$        0.58       & 8.93           $\pm$         0.86        & 9.38           $\pm$         2.28        & 3.93    $\pm$   0.08  \\
G209.00-19.38            & 22.78            &         0.98         & 5.70         $\pm$         0.53       & 7.22           $\pm$         0.25        & 6.42           $\pm$         0.68        & 1.27    $\pm$   0.01  \\
G232.62+00.99            & 23.36            &         0.53         & 4.99          $\pm$         1.72       & 4.90           $\pm$         0.63        & 4.61           $\pm$         1.58        & 3.27    $\pm$   0.06  \\
G211.59+01.05           & 25.19            &          1.76        & 3.20          $\pm$         0.25       & 8.35           $\pm$         0.20        & 10.5         $\pm$         0.78        & 1.21    $\pm$   0.01 \\
\hline
\end{tabular}

\begin{tablenotes}
\footnotesize
\item  Notes. The abundance of  H$_{2}$S are derived from those of H$_{2}$$^{34}$S.
\end{tablenotes}

\end{table}

\clearpage

\section{Abundance ratios.}
\begin{table}[h]
\centering
\caption{The abundance ratios between S-bearing molecules.} \label{table:range}

\begin{tabular}{cccc}

\hline
Source name   &   {[}H$_{2}$S/H$_{2}$CS{]}    &   {[}H$_{2}$S/HCS$^{+}${]}  &     {[}H$_{2}$CS/HCS$^{+}${]}   \\
\hline

G000.67-00.03	&	1.32 	$\pm$	0.12 	&	21.0 	$\pm$	2.18 	&	15.9 	$\pm$	0.86 	\\
G005.88-00.39	&	5.80 	$\pm$	0.14 	&	78.4 	$\pm$	3.29 	&	13.5 	$\pm$	0.55 	\\
G009.62+00.19	&	5.90 	$\pm$	0.17 	&	83.9 	$\pm$	4.06 	&	14.2 	$\pm$	0.62 	\\
G010.47+00.02	&	4.06 	$\pm$	0.07 	&	43.8 	$\pm$	1.48 	&	10.8 	$\pm$	0.33 	\\
G010.62-00.38	&	5.63 	$\pm$	0.05 	&	70.0 	$\pm$	1.20 	&	12.4 	$\pm$	0.19 	\\
G011.49-01.48	&	1.48 	$\pm$	0.66 	&	14.1 	$\pm$	7.01 	&	9.52 	$\pm$	2.47 	\\
G011.91-00.61	&	6.02 	$\pm$	0.27 	&	95.2 	$\pm$	7.20 	&	15.8 	$\pm$	1.10 	\\
G012.80-00.20	&	3.03 	$\pm$	0.06 	&	31.5 	$\pm$	0.66 	&	10.4 	$\pm$	0.13 	\\
G012.88+00.48	&	4.86 	$\pm$	0.08 	&	66.1	$\pm$	1.51 	&	13.6 	$\pm$	0.26 	\\
G012.90-00.26	&	5.91 	$\pm$	0.16 	&	101 	$\pm$	5.68 	&	17.1 	$\pm$	0.90 	\\
G014.33-00.64	&	3.15 	$\pm$	0.16 	&	47.2 	$\pm$	3.27 	&	15.0 	$\pm$	0.77 	\\
G015.03-00.67	&	7.89 	$\pm$	0.78 	&	88.7 	$\pm$	12.5 	&	11.2 	$\pm$	1.25 	\\
G016.58-00.05	&	4.63 	$\pm$	0.16 	&	71.1 	$\pm$	3.62 	&	15.3 	$\pm$	0.64 	\\
G023.00-00.41	&	7.49 	$\pm$	0.32 	&	117 	$\pm$	11.2 	&	15.7 	$\pm$	1.44 	\\
G023.44-00.18	&	3.37 	$\pm$	0.14 	&	46.9 	$\pm$	2.90 	&	13.9 	$\pm$	0.73 	\\
G027.36-00.16	&	3.53 	$\pm$	0.07 	&	73.9 	$\pm$	2.66 	&	20.9 	$\pm$	0.68 	\\
G028.86+00.06	&	6.26 	$\pm$	0.15 	&	76.0 	$\pm$	2.51 	&	12.1 	$\pm$	0.35 	\\
G029.95-00.01	&	8.28 	$\pm$	0.16 	&	118 	$\pm$	5.46 	&	14.2 	$\pm$	0.65 	\\
G031.28+00.06	&	2.97 	$\pm$	0.14 	&	405 	$\pm$	2.62 	&	13.6 	$\pm$	0.63 	\\
G031.58+00.07	&	3.84 	$\pm$	0.19 	&	67.5 	$\pm$	4.43 	&	17.6 	$\pm$	0.87 	\\
G032.04+00.05	&	4.90 	$\pm$	0.12 	&	63.8 	$\pm$	2.63 	&	13.0 	$\pm$	0.47 	\\
G034.39+00.22	&	3.72 	$\pm$	0.30 	&	63.7 	$\pm$	6.78 	&	17.1 	$\pm$	1.34 	\\
G035.02+00.34	&	6.01 	$\pm$	0.13 	&	69.9 	$\pm$	2.21 	&	11.6 	$\pm$	0.30 	\\
G035.19-00.74	&	3.84 	$\pm$	0.17 	&	59.0 	$\pm$	3.37 	&	15.4 	$\pm$	0.59 	\\
G035.20-01.73	&	7.98 	$\pm$	0.37 	&	92.1 	$\pm$	5.72 	&	11.6 	$\pm$	0.59 	\\
G037.43+01.51	&	3.79 	$\pm$	0.18 	&	50.6 	$\pm$	3.51 	&	13.4 	$\pm$	0.78 	\\
G043.16+00.01	&	6.14 	$\pm$	0.10 	&	64.9 	$\pm$	1.53 	&	10.6 	$\pm$	0.21 	\\
G043.79-00.12	&	4.27 	$\pm$	0.16 	&	47.7 	$\pm$	2.14 	&	11.2 	$\pm$	0.33 	\\
G049.48-00.36	&	4.94 	$\pm$	0.07 	&	117 	$\pm$	5.00 	&	23.6 	$\pm$	0.96 	\\
G049.48-00.38	&	2.66 	$\pm$	0.02 	&	57.1 	$\pm$	1.07 	&	21.5 	$\pm$	0.37 	\\
G059.78+00.06	&	4.33 	$\pm$	0.13 	&	46.0 	$\pm$	1.84 	&	10.6 	$\pm$	0.34 	\\
G069.54-00.97	&	4.41 	$\pm$	0.22 	&	59.5 	$\pm$	3.68 	&	13.5 	$\pm$	0.55 	\\
G075.76+00.33	&	4.98 	$\pm$	0.16 	&	64.5 	$\pm$	2.64 	&	12.9 	$\pm$	0.35 	\\
G078.12+03.63	&	3.85 	$\pm$	0.28 	&	52.4 	$\pm$	5.50 	&	13.6 	$\pm$	1.17 	\\
G081.75+00.59	&	4.41 	$\pm$	0.17 	&	50.3 	$\pm$	2.67 	&	11.4 	$\pm$	0.47 	\\
G081.87+00.78	&	3.85 	$\pm$	0.07 	&	65.5 	$\pm$	2.23 	&	17.0 	$\pm$	0.51 	\\
G092.67+03.07	&	2.52 	$\pm$	0.07 	&	36.0 	$\pm$	1.21 	&	14.3 	$\pm$	0.28 	\\
G109.87+02.11	&	10.2 	$\pm$	0.49 	&	130 	$\pm$	11.1 	&	12.7 	$\pm$	0.97 	\\
G111.54+00.77	&	5.80 	$\pm$	0.09 	&	75.6 	$\pm$	1.76 	&	13.0 	$\pm$	0.26 	\\
G121.29+00.65	&	3.67 	$\pm$	0.16 	&	43.0 	$\pm$	2.01 	&	11.7 	$\pm$	0.27 	\\
G123.06-06.30	&	2.52 	$\pm$	0.05 	&	33.9 	$\pm$	0.85 	&	13.5 	$\pm$	0.22 	\\
G133.94+01.06	&	3.94 	$\pm$	0.07 	&	49.8 	$\pm$	1.00 	&	12.6 	$\pm$	0.15 	\\
G168.06+00.82	&	10.2 	$\pm$	3.74 	&	108 	$\pm$	52.6 	&	10.6 	$\pm$	3.88 	\\
G176.51+00.20	&	5.03 	$\pm$	0.67 	&	61.3	$\pm$	10.1 	&	12.2 	$\pm$	1.34 	\\
G183.72-03.66	&	3.27 	$\pm$	0.28 	&	35.2 	$\pm$	3.63 	&	10.8 	$\pm$	0.72 	\\
G188.94+00.88	&	4.41 	$\pm$	0.22 	&	47.1 	$\pm$	2.87 	&	10.7 	$\pm$	0.44 	\\
G192.60-00.04	&	7.94 	$\pm$	0.19 	&	72.8 	$\pm$	2.61 	&	9.16 	$\pm$	0.32 	\\
G209.00-19.38	&	8.44 	$\pm$	1.40 	&	80.4 	$\pm$	3.87 	&	9.53 	$\pm$	1.60 	\\
G232.62+00.99	&	3.83 	$\pm$	0.31 	&	30.4 	$\pm$	3.27 	&	7.94 	$\pm$	0.61 	\\
G211.59+01.05	&	6.14 	$\pm$	0.25 	&	61.1 	$\pm$	3.87 	&	9.95 	$\pm$	0.58 	\\

\hline
\end{tabular}
\end{table}

\end{appendix}

\end{document}